\documentclass[]{aastex631}

\usepackage[section]{placeins}

\newcommand{\rcom}[1]{\textcolor{black}{#1}}

\shorttitle{Possible Epoch of Cometary Activity by Chiron}
\shortauthors{Dobson et al.}

\newcommand{\fink}{{\sc Fink}}
\newcommand{\calviacat}{{\sc calviacat }}
\graphicspath{{./}{figures/}}

\begin{document}

\title{The Discovery and Evolution of a Possible New Epoch of Cometary Activity by the Centaur (2060) Chiron}

\author[0000-0002-1105-7980]{Matthew M. Dobson}
\affiliation{Astrophysics Research Centre, School of Mathematics and Physics, Queen's University Belfast, Belfast BT7 1NN, UK}

\author[0000-0003-4365-1455]{Megan E. Schwamb}
\affiliation{Astrophysics Research Centre, School of Mathematics and Physics, Queen's University Belfast, Belfast BT7 1NN, UK}

\author[0000-0003-0250-9911]{Alan Fitzsimmons}
\affiliation{Astrophysics Research Centre, School of Mathematics and Physics, Queen's University Belfast, Belfast BT7 1NN, UK}

\author[0000-0003-1800-8521]{Charles Schambeau}
\affiliation{Florida Space Institute, University of Central Florida, 12354 Research Parkway, Partnership 1 Building, Orlando, FL 32828, USA}

\author[0000-0002-5780-7062]{Aren Beck}
\affiliation{Florida Space Institute, University of Central Florida, 12354 Research Parkway, Partnership 1 Building, Orlando, FL 32828, USA}

\author[0000-0002-7034-148X]{Larry Denneau}
\affiliation{Institute for Astronomy, University of Hawaii at Manoa, Honolulu, HI 96822, USA}

\author[0000-0002-9986-3898]{Nicolas Erasmus}
\affiliation{South African Astronomical Observatory, Cape Town 7925, South Africa}

\author[0000-0003-3313-4921]{A. N. Heinze}
\affiliation{Department of Astronomy and the DIRAC Institute, University of Washington, 3910 15th Ave NE, Seattle, WA 98195, USA}

\author[0000-0002-5738-1612]{Luke J. Shingles}
\affiliation{GSI Helmholtzzentrum f\"{u}r Schwerionenforschung, Planckstraße 1, 64291 Darmstadt, Germany}
\affiliation{Astrophysics Research Centre, School of Mathematics and Physics, Queen's University Belfast, Belfast BT7 1NN, UK}

\author[0000-0001-5016-3359]{Robert J. Siverd}
\affiliation{Institute for Astronomy, University of Hawaii at Manoa, Honolulu, HI 96822, USA}

\author[0000-0001-9535-3199]{Ken W. Smith}
\affiliation{Astrophysics Research Centre, School of Mathematics and Physics, Queen's University Belfast, Belfast BT7 1NN, UK}

\author[0000-0003-2858-9657]{John L. Tonry}
\affiliation{Institute for Astronomy, University of Hawaii at Manoa, Honolulu, HI 96822, USA}

\author[0000-0003-1847-9008]{Henry Weiland}
\affiliation{Institute for Astronomy, University of Hawaii at Manoa, Honolulu, HI 96822, USA}

\author[0000-0002-1229-2499]{David. R. Young}
\affiliation{Astrophysics Research Centre, School of Mathematics and Physics, Queen's University Belfast, Belfast BT7 1NN, UK}

\author[0000-0002-6702-7676]{Michael S. P. Kelley}
\affiliation{Department of Astronomy, University of Maryland, College Park, MD 20742-0001, USA}

\author[0000-0002-3818-7769]{Tim Lister}
\affiliation{Las Cumbres Observatory, 6740 Cortona Drive Suite 102, Goleta, CA 93117, USA}

\author[0000-0003-0743-9422]{Pedro H. Bernardinelli}
\affiliation{Department of Astronomy and the DIRAC Institute, University of Washington, 3910 15th Ave NE, Seattle, WA 98195, USA}

\author[0000-0002-0535-652X]{Marin Ferrais}
\affiliation{Arecibo Observatory, University of Central Florida, HC-3 Box 53995, Arecibo, PR 00612, USA}

\author[0000-0001-8923-488X]{Emmanuel Jehin}
\affiliation{Space sciences, Technologies \& Astrophysics Research (STAR) Institute, University of Liège, Belgium}

\author[0000-0002-8418-4809]{Grigori Fedorets}
\affiliation{Finnish Centre for Astronomy with ESO, University of Turku, FI-20014 Turku, Finland}
\affiliation{Department of Physics, P.O. Box 64, FI-00014 University of Helsinki, Finland}
\affiliation{Astrophysics Research Centre, School of Mathematics and Physics, Queen's University Belfast, Belfast BT7 1NN, UK}

\author[0000-0001-8821-5927]{Susan D. Benecchi}
\affiliation{Planetary Science Institute, 1700 East Fort Lowell Rd., Suite 106, Tucson, AZ 85719, USA}

\author[0000-0002-3323-9304]{Anne J. Verbiscer}
\affiliation{Department of Astronomy, University of Virginia, P.O. Box 400325, Charlottesville, VA 22904-4325, USA}

\author[0000-0001-9505-1131]{Joseph Murtagh}
\affiliation{Astrophysics Research Centre, School of Mathematics and Physics, Queen's University Belfast, Belfast BT7 1NN, UK}


\author[0000-0001-5963-5850]{Ren{\'e} Duffard}
\affiliation{Instituto de Astrofisica de Andaluc{\'i}a - CSIC. Glorieta de la Astronom{\'i}a s/n. Granada, Spain.}

\author[0000-0001-5749-1507]{Edward Gomez}
\affil{Las Cumbres Observatory, School of Physics and Astronomy, Cardiff University, Queens Buildings, The Parade, Cardiff CF24 3AA, UK}

\author[0000-0002-1278-5998]{Joey Chatelain}
\affil{Las Cumbres Observatory, 6740 Cortona Drive Suite 102, Goleta, CA 93117, USA}

\author[0000-0002-4439-1539 ]{Sarah Greenstreet}
\affil{Rubin Observatory/NSF's NOIRLab, 950 N. Cherry Ave, Tucson, AZ 85719, USA}
\affil{Department of Astronomy and the DIRAC Institute, University of Washington, 3910 15th Ave NE, Seattle, WA 98195, USA}
\begin{abstract}

Centaurs are small Solar System objects on chaotic orbits in the giant planet region, forming an evolutionary continuum with the Kuiper belt objects and Jupiter-family comets. Some Centaurs are known to exhibit cometary activity, though unlike comets this activity tends not to correlate with heliocentric distance and the mechanism behind it is currently poorly understood. We utilize serendipitous observations from the Asteroid Terrestrial-impact Last Alert System (ATLAS), Zwicky Transient Facility (ZTF), Panoramic Survey Telescope and Rapid Response System (Pan-STARRS), Dark Energy Survey (DES), and Gaia in addition to targeted follow-up observations from the Las Cumbres Observatory, TRAnsiting Planets and PlanetesImals Small Telescope South (TRAPPIST-South), and Gemini North telescope to analyze an unexpected brightening exhibited by the known active Centaur (2060) Chiron in 2021. This is highly indicative of a cometary outburst. As of 2023 February, Chiron has still not returned to its pre-brightening magnitude. We find Chiron’s rotational lightcurve, phase curve effects, and possible high-albedo surface features to be unlikely causes of this observed brightening. We consider the most likely cause to be an epoch of either new or increased cometary activity, though we cannot rule out a possible contribution from Chiron’s reported ring system, such as a collision of as-yet unseen satellites shepherding the rings. We find no evidence for coma in our Gemini or TRAPPIST-South observations, though this does not preclude the possibility that Chiron is exhibiting a coma that is too faint for observation or constrained to the immediate vicinity of the nucleus.


\end{abstract}

\keywords{
}


\section{Introduction} 

The Centaurs are small icy objects residing on dynamically unstable orbits in the giant planet region of the Solar System. Centaurs are thought to originate as primordial Kuiper belt objects (KBOs) on Neptune-crossing orbits which are then scattered by Neptune into the giant planet region \citep[]{1997Sci...276.1670D,1997Icar..127...13L,2008ApJ...687..714V} where they reside on chaotic orbits with an average orbital lifetime of the order of ${\sim}10^{6}-10^{8}$ years \citep[]{1990Natur.348..132H,2003AJ....126.3122T,2007Icar..190..224D,2008ApJ...687..714V,2009Icar..203..155B}. During this time, Centaurs undergo encounters with the giant planets, causing the Centaurs to either impact the giant planets, be ejected out of the Solar System altogether, or become short period comets which include the Jupiter-family comets (JFCs; \citealt[]{2003AJ....126.3122T,2004come.book..193D,2013ApJ...773...22B,2013Icar..226.1138F,2019ApJ...883L..25S}). Centaurs thus form an evolutionary link between the Kuiper belt and the more physically evolved JFCs, and this intermediate transitionary nature makes them valuable in studying the evolution of JFCs.

Approximately 8-10\% of Centaurs are known to exhibit dust comae, indicating cometary activity \citep{2020tnss.book..307P}. 
\rcom{However, for most active Centaurs, especially those with large perihelia, their activity differs from that of JFCs in that it occurs throughout their orbits \citep{2020tnss.book..307P}. Centaur activity appears to be dependent on a Centaur's perihelion distance from the Sun, with a possible boundary existing at ${\sim}14$ au beyond which no activity has been detected. \citep[]{2009AJ....137.4296J,2012AJ....144...97G,2015AJ....150..201J,2020AJ....159..209L,2021PSJ.....2..155L,2024ApJ...960L...8L}.} The mechanism behind this phenomenon currently remains \rcom{unknown}, with newly-exposed pockets of volatile surfaces ices \citep{1992ApJ...388..196P}
as well as volatiles released during the amorphous-to-crystalline transition of sub-surface ice \citep{2009AJ....137.4296J} being proposed to explain Centaur activity.
Known active Centaurs tend to reside in orbits with smaller semimajor axes than inactive ones \citep{2009AJ....137.4296J}
with active Centaurs tending to have experienced recent ($\sim10^{2}$ - $10^{3}$ years) dynamical encounters with the giant planets, causing a significant decrease in their semimajor axes \citep[]{2018P&SS..158....6F,2021PSJ.....2..155L,2024ApJ...960L...8L}. Combined with the comparative lack of planetary encounters experienced by inactive Centaurs, this hints that a sudden decrease in perihelia could trigger the onset of Centaur activity \citep[]{2018P&SS..158....6F,2021PSJ.....2..155L,2024ApJ...960L...8L}.

(2060) 95P/Chiron (henceforth Chiron) is one of the largest members of the Centaur population and has a 
well-established history of cometary activity 
\citep[]{1989Icar...77..223B,1989IAUC.4770....1M,1990Icar...83....1H,1990AJ....100..913L,1990IAUC.4947....3M,1990IAUC.4970....1W,1990AJ....100.1323M,1990nba..meet...83D,1991A&A...241..635W,1991ApJ...366L..47B,1991Sci...251..777S,1991MNRAS.250..115D,1993Icar..104..234M,1993PASP..105..946L,1993IAUC.5898....1B,1995Natur.373...46E,1996Icar..123..478B,1996P&SS...44.1547L,1997P&SS...45.1607L,1988IAUC.4554....2T,2001Icar..150...94B,2001PSS...49.1325S,2010Icar..210..472B}.
Cometary activity was first discovered in 1989 when a coma was detected around Chiron as it approached perihelion 
\citep[]{1989IAUC.4770....1M,1990Icar...83....1H,1990IAUC.4947....3M}
and remained detectable even when the absolute magnitude of Chiron was at its dimmest recorded value near perihelion
\citep{2001PSS...49.1325S}.  Subsequent analysis of archival photographic plates obtained from 1969 to 1977 has also revealed that Chiron was active near aphelion 
\citep{2001Icar..150...94B}. 
This cometary activity is thought to contribute to the considerable variation in the brightness of Chiron observed over time 
\citep[]{2001Icar..150...94B,2002Icar..160...44D,2010Icar..210..472B}. Models of Chiron's nucleus and reported ring system require a contribution from cometary outbursts to reproduce Chiron's measured lightcurve \citep{2015A&A...576A..18O}.

Between 2021 February and 2021 June, Chiron underwent a sudden brightening in apparent magnitude by ${\sim}1$ mag, remaining brighter than previous years for all of its 2021-2022 observing season \citep[]{2021RNAAS...5..211D,2023PSJ.....4...75D,2023MNRAS.523.3678B,2023arXiv230803458O}, potentially indicative of an epoch of either new or increased cometary activity \citep[]{2021RNAAS...5..211D,2023PSJ.....4...75D,2023arXiv230803458O}. We refer to this observed brightening as the 2021 Brightening Event hereafter. 
In this paper, we analyze the 2021 Brightening Event exhibited by Chiron. We utilise multi-wavelength photometry from the Asteroid Terrestrial-impact Last Alert System (ATLAS, \citealt[]{2018PASP..130f4505T,2018ApJ...867..105T}), the Zwicky Transient Facility (ZTF; \citealt[]{2019PASP..131a8002B,2019PASP..131g8001G}), and the Panoramic Survey Telescope and Rapid Response System (Pan-STARRS; \citealt[]{2002SPIE.4836..154K,2004SPIE.5489...11K,2004AN....325..636H,2013PASP..125..357D,2016arXiv161205560C,2020ApJS..251....3M,2020ApJS..251....4W,2020ApJS..251....5M,2020ApJS..251....6M,2020ApJS..251....7F}) spanning multiple observing seasons, in addition to follow-up photometric observations taken by the 8.1-m telescope of the Gemini Observatory \citep[]{2004PASP..116..425H,2016SPIE.9908E..2SG} and the Las Cumbres Observatory (LCO) telescope network \citep{2013PASP..125.1031B} via the LCO Outbursting Objects Key (LOOK) project \citep{2022PSJ.....3..173L}. We further supplement our dataset with observations from the Gaia Data Release 3 (DR3, \citealt[]{2023A&A...674A...1G,2023A&A...674A..12T}), the robotic TRAnsiting Planets and PlanetesImals Small Telescope South (TRAPPIST-South, \citealt{2011Msngr.145....2J}), the Dark Energy Survey (DES, \citealt{2005astro.ph.10346T}), and  observations of Chiron reported in the literature.
Our paper is structured as follows. In Section \ref{ObservationsAndDataReduction}, we present details of the photometry used in this study and the data reduction techniques used for each dataset. We present our methods of analysing our observations in Section \ref{DataAnalysis}. In Section \ref{Results}, we present our results of analysing the 2021 Brightening Event for photometric behaviour, color change, search for coma, and determining contribution from Chiron's reported ring system. We discuss the results of our analysis and their implications for the cause of the 2021 Brightening Event in Section \ref{Discussion}. We list the conclusions of our study in Section \ref{Conclusions}.

\section{Observations and Data Reduction} \label{ObservationsAndDataReduction}


This Section briefly summarises each telescope and/or survey and describes the observations we use in our analysis. Our dataset is comprised of targeted observations (TRAPPIST-South, Gemini, and LOOK) and survey observations (Pan-STARRS, DES, Gaia, ATLAS, and ZTF) where Chiron was serendipitously imaged. Our observations range in time from 2012 to 2023, sampling Chiron's lightcurve pre-, during, and post-2021 Brightening Event. 
Tables of all observations from each telescope/survey are listed in Appendix \ref{Appendix}, with the number of observations in each filter from each telescope/survey listed in Table \ref{ChironSurveyDatasets}. Most of our observations are from the ATLAS dataset, followed by ZTF, and combined, ATLAS and ZTF cover the largest timespan in our sample. These large datasets therefore constitute the majority of our analysis. Pan-STARRS provides our earliest consistent observations of Chiron, extending our baseline to 2009. Our observations from Gemini occurred just after the 2021 Brightening Event and allow us to search for any coma exhibited by Chiron during this time.
Our observations span 12 observing seasons when Chiron's on-sky distance from the Sun was sufficient to allow observations.
The date ranges of each observing season are listed in Table \ref{ChironObservingSeasons}. For clarity, we designate each observing season with a letter as described in Table \ref{ChironObservingSeasons}.

\begin{deluxetable*}{lcccr}[h] \label{ChironSurveyDatasets}
\tablecaption{Number and Time Ranges of Observations of Chiron for Each Filter of Each Telescope/Survey}
\tablecolumns{4}
\tablehead{
\colhead{Telescope/Survey} & 
\colhead{Filter} & 
\colhead{Start Date} & 
\colhead{End Date} &
\colhead{Number of Observations}\\
\colhead{} & 
\colhead{} & 
\colhead{(YYYY-MM-DD)} &
\colhead{(YYYY-MM-DD)} &
\colhead{}
}
\startdata
     Pan-STARRS &        g &    2010-09-05 &    2013-08-07 &                       8 \\
     Pan-STARRS &        r &   2010-08-31 &   2013-08-17 &                      11 \\
     Pan-STARRS &        i &    2010-09-02 &  2014-10-11 &                      12 \\
     Pan-STARRS &        z &   2010-06-22 &    2014-07-03 &                      12 \\
     Pan-STARRS &        y &   2009-06-10 &  2013-10-30 &                      12 \\
 TRAPPIST-South &        B &   2012-05-16 &   2012-09-12 &                       8 \\
 TRAPPIST-South &        V &   2012-05-16 &   2012-09-12 &                       9 \\
 TRAPPIST-South &        R &   2012-05-16 &   2015-09-18 &                      82 \\
            DES &    DES-g &   2013-09-11 &   2016-10-05 &                       5 \\
            DES &    DES-r &  2015-10-31 &   2016-10-05 &                       3 \\
            DES &    DES-i &  2015-10-31 &   2017-11-08 &                       4 \\
            DES &    DES-z &   2016-09-30 &  2017-11-24 &                       4 \\
            DES &    DES-Y &  2014-10-28 &   2016-11-04 &                       3 \\
           Gaia &   Gaia-G &  2014-10-31 &   2017-05-24 &                     155 \\
          ATLAS &        c &   2015-08-12 &  2022-12-17 &                     399 \\
          ATLAS &        o &   2015-07-26 &   2023-01-20 &                    1369 \\
            ZTF &    ZTF-g &   2019-11-02 &    2023-01-07 &                      89 \\
            ZTF &    ZTF-r &   2019-11-02 &   2023-01-12 &                     103 \\
         Gemini &  SDSS-g' &   2021-08-20 &    2022-08-02 &                       5 \\
         Gemini &  SDSS-r' &   2021-08-20 &    2022-08-02 &                       3 \\
         Gemini &  SDSS-i' &   2021-08-20 &    2022-08-02 &                       4 \\
           LOOK &   LOOK w &    2021-09-06 &    2021-09-06 &                       4 \\
           LOOK &  SDSS-g' &    2022-09-04 &    2023-02-05 &                      22 \\
           LOOK &  SDSS-r' &    2022-09-04 &    2023-02-05 &                      21 \\
\enddata
\end{deluxetable*}

\begin{deluxetable*}{lccr}[h] \label{ChironObservingSeasons}
\tablecaption{Date Ranges and Survey Coverage of Each Observing Season of Chiron}
\tablecolumns{4}
\tablehead{
\colhead{Observing} & 
\colhead{UT Start Date} & 
\colhead{UT End Date} & 
\colhead{Telescope/Survey} \\
\colhead{Season} &
\colhead{of Data Coverage} &
\colhead{of Data Coverage} &
\colhead{Data Coverage} \\
\colhead{} &
\colhead{(YYYY-MM-DD)} &
\colhead{(YYYY-MM-DD)} &
\colhead{}
}
\startdata
A & 2009-06-10 & 2009-06-10 & Pan-STARRS\\
B & 2010-05-29 & 2010-11-02 & Pan-STARRS\\
C & 2011-06-17 & 2011-10-13 & Pan-STARRS\\
D & 2012-05-16 & 2012-11-16 & Pan-STARRS, TRAPPIST-South \\
E & 2013-05-16 & 2013-10-30 & DES, Pan-STARRS, TRAPPIST-South\\
F & 2014-07-03 & 2014-12-19 & DES, Gaia, Pan-STARRS, TRAPPIST-South\\
G & 2015-05-03 & 2015-12-30 & ATLAS, DES, Gaia, TRAPPIST-South\\
H & 2016-05-13 & 2017-01-09 & ATLAS, Gaia, DES\\
I & 2017-05-24 & 2018-01-17 & ATLAS, Gaia, DES\\
J & 2018-06-28 & 2019-01-24 & ATLAS \\
K & 2019-05-26 & 2020-01-31 & ATLAS, ZTF\\
L & 2020-05-19 & 2021-02-06 & ATLAS, ZTF\\
M & 2021-06-17 & 2022-01-20 & ATLAS, Gemini, ZTF\\
N & 2022-06-24 & 2023-02-05 & ATLAS, Gemini, LOOK, ZTF\\
\enddata
\tablecomments{Observing seasons A-L are before the 2021 Brightening Event, which occured during observing season M. Observing season N is after the 2021 Brightening Event}
\end{deluxetable*}

\subsection{ATLAS}


ATLAS consists of four 0.5-m Schmidt telescopes at sites in Hawai'i, Chile, and South Africa, each covering a 28.9 deg$^{2}$ field-of-view (FOV) per exposure \citep[]{2018PASP..130f4505T,2018ApJ...867..105T}. ATLAS regularly observes the sky to a limiting magnitude of ${\sim}$19.5 mag in two non-standard wide-band filters, cyan ($c$, spanning 420-650 nm) and orange ($o$, spanning 560-820 nm). 
Further details of the ATLAS system and data reduction pipeline are described in \citet[]{2018PASP..130f4505T,2018ApJ...867..105T} and \citet{2020PASP..132h5002S}. 
We utilise serendipitous observations from all four ATLAS telescope sites, obtaining our magnitude measurements of Chiron via the ATLAS Forced Photometry Server\footnote{\url{https://fallingstar-data.com/forcedphot/}} \citep{2021TNSAN...7....1S}, which fits a point spread function (PSF) at a given object's position on the image as predicted from its orbital ephemeris catalogued by the Minor Planet Center, and calculates the AB magnitude of the flux at that point. 
We chose to use difference images generated from the ATLAS data reduction pipeline \citep[]{2018PASP..130f4505T,2018ApJ...867..105T} instead of reduced images to reduce the effect of contaminating background stars and galaxies on the brightness measurements of Chiron. 
We select all ATLAS data up to and including 2023 January 20, corresponding to the most recent observation of Chiron's observing season from 2022-2023. From these, we select for analysis all ATLAS observations with apparent magnitudes brighter than both (i) the $5\sigma$ limiting magnitude of the image and (ii) the $3\sigma$ upper magnitude limit derived from the flux uncertainty, to ensure all measurements are from good quality observations. All ATLAS data used for our analysis extend from 2015 July 26 to 2023 January 20. The majority of these data came from the telescopes at the two Hawai'ian sites, with contributions from all four telescopes from 2022 July 31 onwards. All ATLAS data used for our analysis are listed in Table \ref{ATLASTable}.

\subsection{ZTF}



The Zwicky Transient Facility (ZTF; \citealt[]{2019PASP..131a8002B,2019PASP..131g8001G}) is a wide-field time-domain sky survey \citep{2019PASP..131g8001G} utilizing the 48-inch (1.2-m) Samuel Oschin Schmidt Telescope at the Palomar Observatory with a 43 deg$^{2}$ FOV. Observations have been collected since 2017 October to a limiting magnitude of ${\sim}20.5$ mag with 30 second exposures. 
Further details of ZTF, its setup and survey strategy are described in \citet{2019PASP..131a8002B} and \citet{2019PASP..131g8001G}.
ZTF has accumulated photometry of Chiron in $ZTF{-}g$ and $ZTF{-}r$ band, extending in time from 2019 November 2 to 2023 January 12. We obtain magnitude measurements from these serendipitous observations using the \fink\footnote{\url{https://fink-portal.org}} community broker \citep{2021MNRAS.501.3272M}, which processes all alerts of detections by ZTF and links them to the corresponding known Solar System objects. Table \ref{ZTFTable} lists the magnitude values obtained via PSF photometry for all ZTF observations of Chiron used in this work.

\subsection{Pan-STARRS}

The Panoramic Survey Telescope and Rapid Response System \citep[]{2002SPIE.4836..154K,2004SPIE.5489...11K,2004AN....325..636H,2013PASP..125..357D,2016arXiv161205560C,2020ApJS..251....3M,2020ApJS..251....4W,2020ApJS..251....5M,2020ApJS..251....6M,2020ApJS..251....7F} consists of two 1.8-m telescopes located at Haleakala Observatory, Hawai'i, observing the entire visible sky in six broadband filters, $g$, $r$, $i$, $z$, $y$, and $w$ to a limiting magnitude of $r\sim21.2$ mag. Each telescope has a ${\sim}$7 deg field of view and is equipped with a 1.4 gigapixel camera, resulting in a 0.26 arcsec/pixel resolution \citep{2013PASP..125..357D}. Further details of Pan-STARRS, its design, survey strategy, and data analysis are descibed in \citet[]{2016arXiv161205560C,2020ApJS..251....3M,2020ApJS..251....4W,2020ApJS..251....5M,2020ApJS..251....6M}; and \citet{2020ApJS..251....7F}.
Throughout its observation baseline, Pan-STARRS has obtained serendipitous observations of Chiron in the $g$, $r$, $i$, $z$, and $y$ filters, ranging in time from 2009 June 10 to 2014 October 11. We utilize the Canadian Astronomy Data Center (CADC) Solar System Object Image Search\footnote{\url{https://www.cadc-ccda.hia-iha.nrc-cnrc.gc.ca/en/ssois/}} \citep{2012PASP..124..579G} to obtain magnitude measurements of Chiron from these observations, utilizing the Catalog Archive Server Jobs System\footnote{\url{http://casjobs.sdss.org/CasJobs}} (CasJobs), developed by the Johns Hopkins University/Sloan Digital Sky Survey (JHU/SDSS) team, to query the photometry. Table \ref{PanSTARRSTable} lists all the Pan-STARRS magnitude values from PSF photometry of Chiron used in this work. 

\subsection{Dark Energy Survey}

We use the public images from DES \citep{2005astro.ph.10346T} Data Release 2 \citep{2021ApJS..255...20A}. DES is an imaging survey
utilizing the 570 MPix Dark Energy Camera (DECam, \citealt{2015AJ....150..150F})
of the 4-m Blanco telescope at the Cerro Tololo Interamerican
Observatory (CTIO). DES serendipitously observed Chiron during its operations between 2013-2019 
\citep[]{2020ApJS..247...32B,2022ApJS..258...41B,2022AAS...24033207B,2023ApJS..269...18B}. The Chiron images used in this work were
processed using the scene modeling photometry method of \citet{2023ApJS..269...18B}. In addition to the photometric uncertainty derived from the scene-modeling procedure, an additional 3 mmag uncertainty was added in quadrature to account for the known scatter in the DES calibration with respect to Gaia \citep{2021ApJS..255...20A}. 
The resulting magnitude values from PSF photometry are listed in Table \ref{DESTable}.

\subsection{Gaia}


We include in our analysis serendipitous observations of Chiron from \textit{Gaia} Data Release 3 (DR3, \citealt[]{2016A&A...595A...1G,2018A&A...616A..13G,2023A&A...674A..12T,2023A&A...674A...1G}) in Table \ref{GaiaTable}. These observations obtained from the space-based \textit{Gaia} telescope span from 2014 October 31 to 2017 May 24 in the visible white-light G-band filter (wavelength range 330-1050 nm, \citealt{2010A&A...523A..48J}). 

\subsection{LOOK}

The Las Cumbres Observatory (LCO) operates 25 telescopes across the globe \citep{2013PASP..125.1031B}.
As part of the LCO Outbursting Objects Key Project (LOOK; \citealt{2022PSJ.....3..173L})
we observed Chiron from 2022 September 18 to 2023 February 6 using the thirteen 1-m telescopes of the LCO network in Pan-STARRS $g'$ and $r'$ filters \citep{2016arXiv161205560C}, utilizing the NEOExchange web portal \citep{2021Icar..36414387L} to schedule our targeted observations. Each 1-m telescope utilizes a Sinistro imager comprising a Fairchild $4096\times4096$ pixel CCD and an Archon controller, giving a field of view of $26.6' \times 26.6'$ with a resolution of 0.387 arcsec/pixel \citep{2013PASP..125.1031B}. We also include the targeted observations from \citet{2021RNAAS...5..211D} taken in the $w$ band 
(equivalent to $g'$+$r'$+$i'$ band) on 2021 September 6. All observations were obtained with the telescope in half-rate tracking mode, with exposure times ranging from 180-245 seconds ensuring negligible on-sky motion of Chiron. These observations are automatically processed using the LCO image processing pipeline ``Beautiful Algorithm to Normalize Zillions of Astronomical Images" (BANZAI; \citealt{2018SPIE10707E..0KM}) and automatically calibrated to the Pan-STARRS1 photometric system \citep{2012ApJ...750...99T} using the ATLAS-RefCat2 all-sky photometric catalog \citep{2018ApJ...867..105T}, the \calviacat software \citep{2021zndo...5061298K}, and background field stars measured with BANZAI \citep{2022PSJ.....3..173L}. We apply aperture photometry to the processed images utilizing a 5 arcsec aperture radius to all observations with seeing ${\leq}5$ arcsec. The resulting magnitude values of each targeted observation and their associated conditions are listed in Table \ref{LOOKTable}. 

\subsection{TRAPPIST-South}

We also include targeted observations of Chiron acquired between 2012 and 2015 with TRAPPIST-South \citep{2011Msngr.145....2J} located at the La Silla Observatory in Chile at the European Southern Observatory (ESO). TRAPPIST-South is a 0.6-m Ritchey-Chrétien telescope with a thermoelectrically-cooled FLI ProLine 3041-BB CCD camera providing a field of view of $22' \times 22'$ and a sampling of 1.2 arcsec/pixel in the 2x2 \rcom{binned} mode. The data were reduced using the PHOTOMETRYPIPELINE \citep{2017A&C....18...47M} with photometric apertures with radii of 4 pixels and calibrated to the band in which they were observed (Johnson-Cousins $B$, $V$, and $R$) using the Pan-STARRS catalog \citep[]{2016arXiv161205560C,2020ApJS..251....3M,
2020ApJS..251....7F}. Table \ref{TRAPPISTSouthTable} lists all the data from TRAPPIST-South used for our analysis.


\subsection{Gemini}

We obtained $g$, $r$, and $i$ band photometry of Chiron via targeted observations taken with the Gemini Observatory from the program GN-2021B-FT-114 and through observations acquired through the Gemini Large and Long Program (LLP) ``Investigating the Activity Drivers of Small Bodies in the Centaur-to-Jupiter-Family Transition" (programs GN-2021B-LP-203, GN-2022A-LP-203, and GN-2022B-LP-203).
Chiron data were obtained from the 8.1-m Gemini North telescope in Hawai'i using the Gemini Multi-Object Spectrograph (GMOS, \citealt[]{2004PASP..116..425H,2016SPIE.9908E..2SG}) in imaging mode. GMOS-N has a field of view of $5.5' \times 5.5'$ with a pixel resolution of 0.1614 arcsec/pixel with 2x2 binning. Observations were made in the $g$, $r$, and $i$ bands at five epochs (2021 August 8, 2021 November 3, 2021 December 9, 2022 June 24, and 2022 August 2) with all observations sidereally tracked. The exposure times of each individual observation in a sequence were 
short enough that Chiron’s on-sky motion (maximum 5.53 arcsec/hr) would be negligible at the 2x2 binned pixel scale, ensuring a symmetric PSF.
Table \ref{tab:LLP-obs} provides details of the observing circumstances for all our observing sequences of Chiron with Gemini.
All Gemini GMOS images were reduced, including bias subtraction, flat-field correction, and mosaicing of the three charge-coupled device (CCD) images, using the Gemini Data Reduction for Astronomy from Gemini Observatory North and South ({DRAGONS}) Python package \citep[]{2019ASPC..523..321L,Labrie_2023}. 
Individual images were then processed using our custom Python-based GMOS reduction and calibration pipeline.
The pipeline includes: (1) cosmic-ray removal performed utilizing the LACosmic technique \citep{van-dokkum-2001} as implemented in ccdproc \citep[]{craig-2017,curtis_mccully_2018_1482019}, (2) stacking of individual images in Chiron's non-sidereal frame and the sidereal frame, and (3) a Pan-STARRS-based calibration using field stars from the GMOS field of view stacked in the sidereal frame, utilizing their catalog magnitudes as archived in the PS1 MeanObject tables from the Pan-STARRS Data Release 2 \citep[]{2016arXiv161205560C,2020ApJS..251....3M,
2020ApJS..251....7F}. We utilized Source Extractor Python \citep[]{1996A&AS..117..393B,2016JOSS....1...58B} to detect background stellar sources in our Gemini observations.

\section{Data Analysis} \label{DataAnalysis}

For all our observations, we correct the time of observation for light travel time and calculate reduced magnitudes (the apparent magnitude scaled to a geocentric distance and heliocentric distance of 1 au) and their associated uncertainties, assuming negligible contribution from heliocentric and geocentric distances (obtained from JPL Horizons\footnote{\url{https://ssd.jpl.nasa.gov}}) to the errors. To help remove any stellar contamination or spurious datapoints from our analysis we split our observations into pre- and during/post-2021 Brightening Event. We treat each filter of each dataset individually. 
For datasets with ${<}50$ observations - namely, DES, Gemini, Pan-STARRS, and LOOK in all filters, and TRAPPIST-South in $B$ and $V$ filters - we visually inspect each image, removing any observations which exhibit evidence of contamination from background sources. To remove possibly spurious datapoints for datasets with $\geq50$ observations, we followed the prescription \citet{2023PSJ.....4...75D} of removing datapoints with large magnitude uncertainties and $\sigma$-clipping our dataset from. 

All observations of Chiron pre-2021 Brightening Event, corresponding to observing seasons A-L (2009-2021), are consistent with a single phase curve slope. Analysing ATLAS and ZTF observations, which constitute the majority of our data and span observing seasons G-L (2015-2021), utilizing the same method as \citet{2023PSJ.....4...75D}, we find that their reduced magnitude values in a given filter all reside on a single phase curve profile (see Figure \ref{ChironATLASLightCurvePhaseCurvesObservingSeasonsCyanOrange} in Section \ref{BrightnessEvolution} for more details). Therefore, we treat all observing seasons pre-2021 Brightening Event as one. From this time range, we remove all observations with magnitude uncertainties exceeding the 85th percentile of the magnitude uncertainty distribution for ATLAS and TRAPPIST-South in the $R$ filter and 95th percentile for ZTF and Gaia. For each filter in our ATLAS, ZTF, TRAPPIST-South $R$, and Gaia datasets, we $3\sigma$-clip the remaining observations to eliminate outliers in reduced magnitude. 

For all observations post-2021 Brightening Event, we treat each observing season separately due to Chiron's evolving phase curve (see Section \ref{Results} for more details). We utilise the 95th percentile of the magnitude uncertainty distribution as the cut-off for our ATLAS and ZTF datasets. For a given observing season, we bin all remaining observations into two bins of phase angle, each of width equal to half the maximum phase angle span of the dataset. This ensures that the algorithm does not remove features of the phase curve (e.g. opposition effect) which may not be fully sampled by the dataset of one observing season. For each observing season, we $3\sigma$-clip the observations in each bin. The comparatively smaller sizes of the Gemini and LOOK datasets means we instead manually remove any outliers caused by contamination from background sources in the images.

\section{Results} \label{Results}

In this section, we analyse the 2021 Brightening Event and its subsequent evolution to explore the potential causes of Chiron’s enhanced brightness in all filters. There are several possible explanations for the 2021 Brightening Event that we test with our dataset. 
Cometary activity could cause Chiron to increase in brightness. Chiron is a known active object with previous instances of the Centaur brightening coinciding with detections of visible coma and/or significant PSF extension \citep[]{1989Icar...77..223B,1989IAUC.4770....1M,1990Icar...83....1H,1990AJ....100..913L,1990IAUC.4947....3M,1990IAUC.4970....1W,1990AJ....100.1323M,1990nba..meet...83D,1988IAUC.4554....2T,2001Icar..150...94B,2001PSS...49.1325S}. Alternatively, Chiron’s ring system may instead offer an explanation for the 2021 Brightening Event. Occultation studies have revealed Chiron to be surrounded by either symmetric shell-like structures \citep{2015Icar..252..271R} or a system of debris rings \citep[]{2015A&A...576A..18O,2023arXiv230803458O}. \citet{2015A&A...576A..18O} found that by modeling the contribution of rings to the Centaur’s absolute magnitude across its orbit, their changing orientation as viewed from Earth could explain previous instances of brightening by Chiron. \citet{2023MNRAS.523.3678B} hypothesised that the 2021 Brightening Event could instead be caused by variation in Chiron’s surface albedo, with the seasonal appearance of a high-albedo surface feature causing the Centaur to brighten. We explore the brightness evolution and phase curve of Chiron pre-, during, and post-2021 Brightening Event (Section \ref{BrightnessEvolution}); search for significant changes in color index across time (Section \ref{ColourEvolution}); search for visible coma or PSF extension in our observations (Section \ref{ComaSearch}); and compute the effect of Chiron's ring system on its brightness evolution across our baseline of observations (Section \ref{Rings}).

\subsection{Brightness Evolution} \label{BrightnessEvolution}

\begin{figure}
\centering
\includegraphics[height=0.95\textheight]{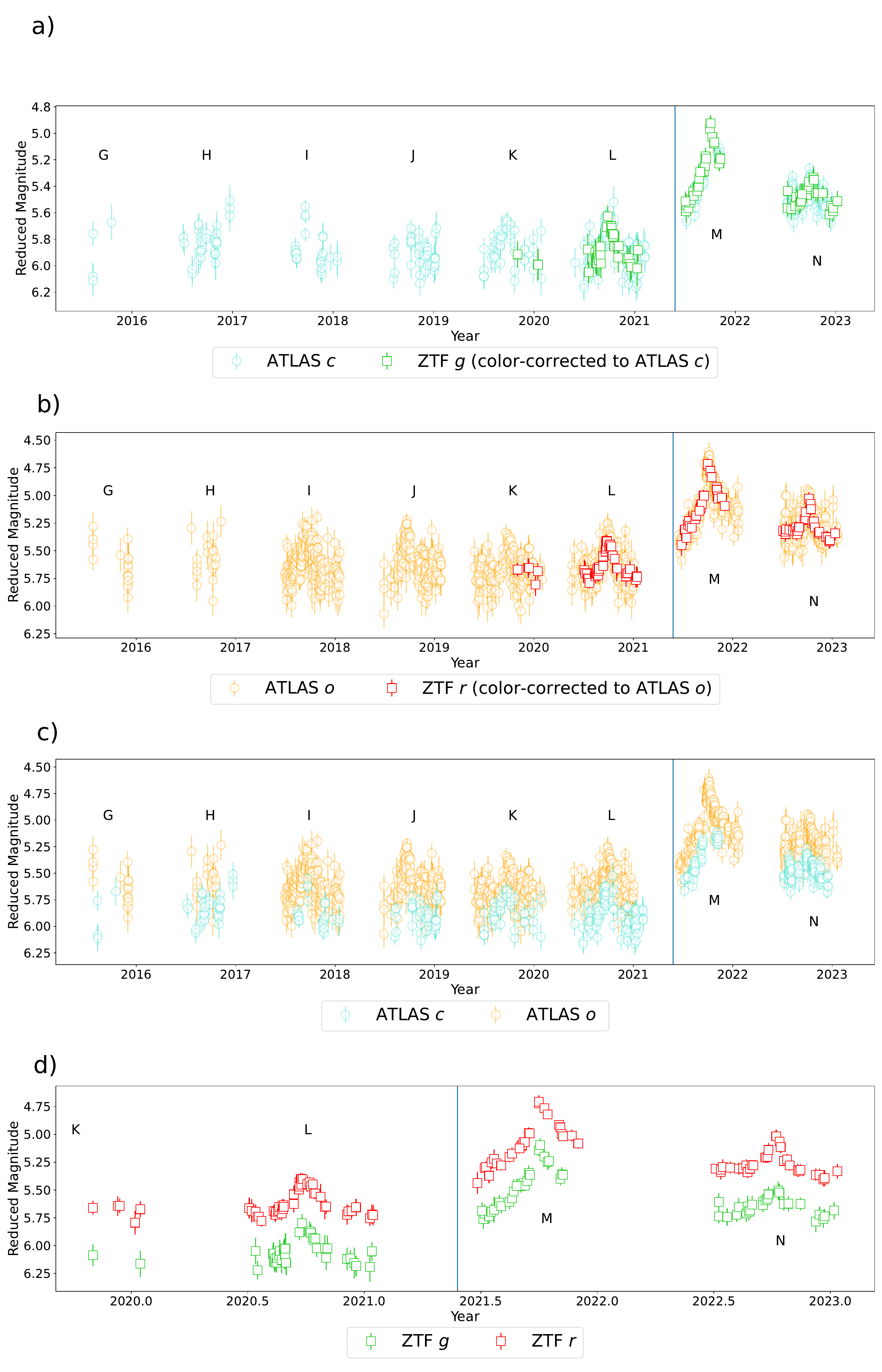}
\caption{\rcom{a) Color-corrected reduced magnitudes vs. time of Chiron in ATLAS c and ZTF g filters. b) Color-corrected reduced magnitudes vs. time of Chiron in ATLAS o and ZTF r filters. c) ATLAS reduced magnitude values vs. time. d) ZTF reduced magnitude values vs time. Error bars are $1\sigma$ uncertainties. Note that the ATLAS and ZTF reduced magnitudes are not color-corrected in this figure. All observing seasons are labeled as per Table \ref{ChironObservingSeasons}. The earliest data in this plot start at observing season G (2015-2016) as observing seasons A-F (2012-2014) have insufficient data to fully sample Chiron's brightness across time. Vertical blue lines indicates time at which Chiron was at aphelion.
}}
\label{ChironDataLightcurve}
\end{figure}

\begin{figure}
\centering
\includegraphics[height=0.95\textheight]{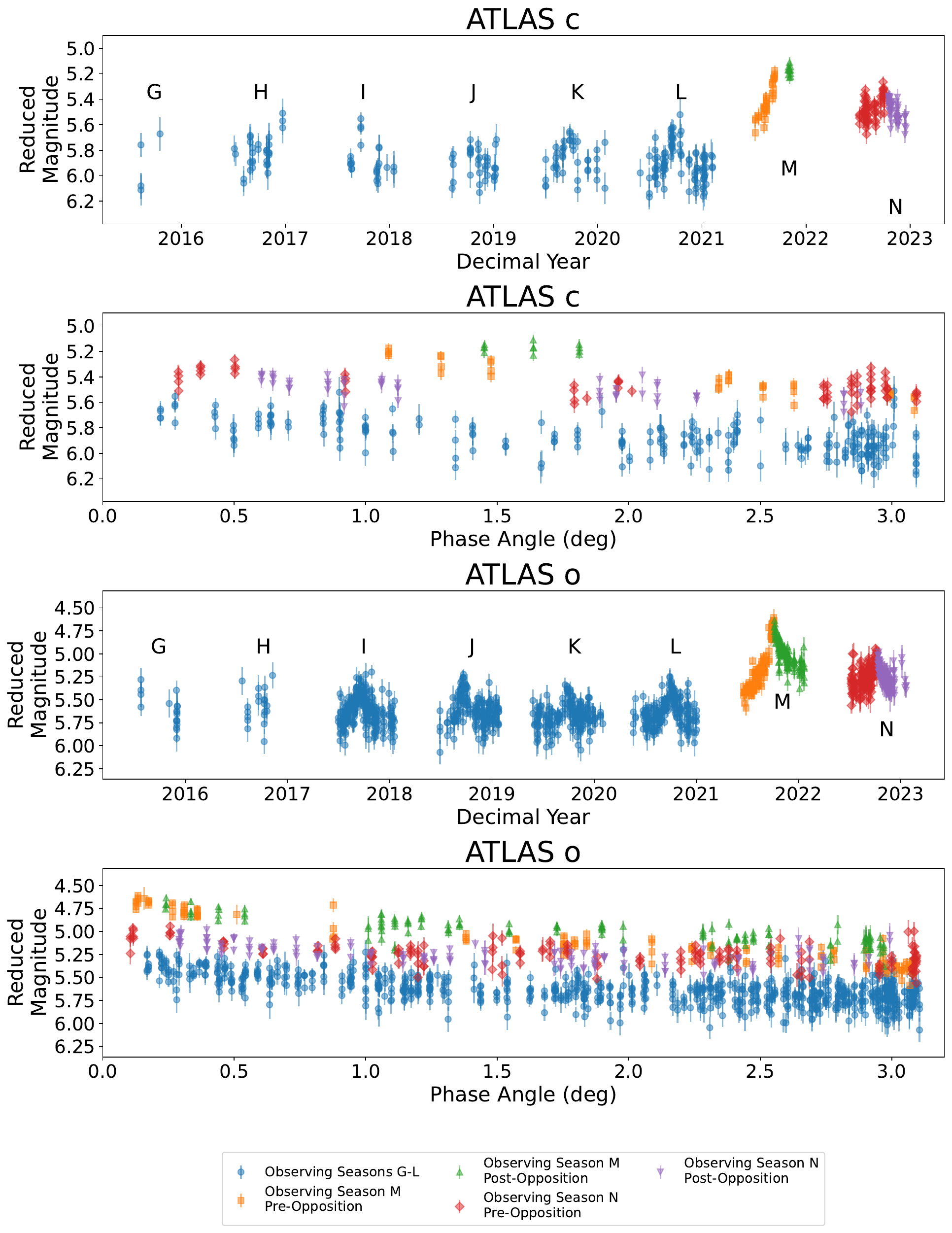}
\caption{\textit{(a):} ATLAS $c$ filter lightcurve of Chiron. \textit{(b):} ATLAS $c$ filter phase curve of Chiron. \textit{(c):} ATLAS $o$ filter lightcurve of Chiron. \textit{(d):} ATLAS $o$ filter phase curve of Chiron. $1\sigma$ uncertainty error bars are plotted though may be smaller than plot markers. \rcom{All observing seasons are labeled as per Table \ref{ChironObservingSeasons}.}
}
\label{ChironATLASLightCurveCyanOrange}
\end{figure}

\begin{figure}
\centering
\includegraphics[height=0.95\textheight]{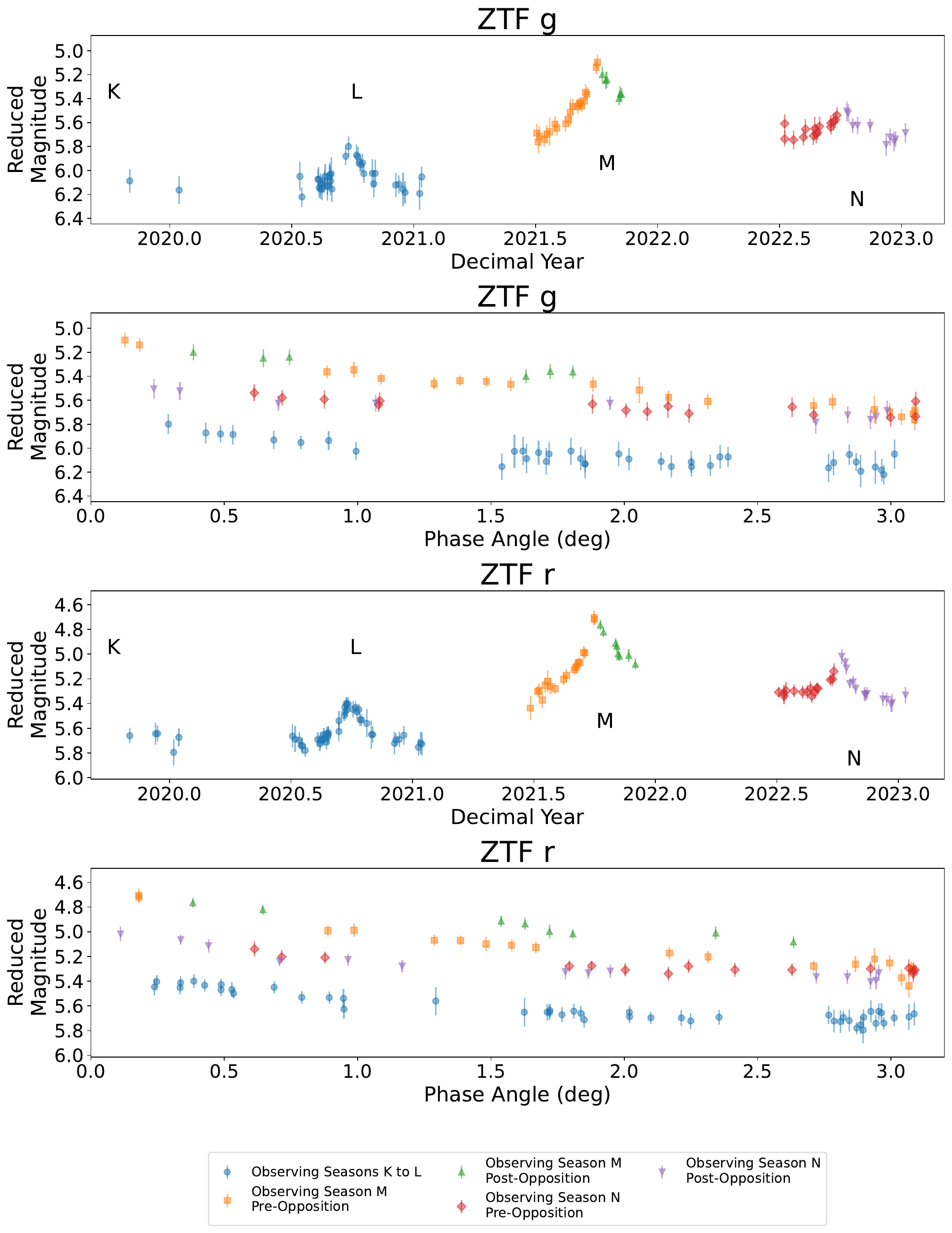}
\caption{\textit{(a):} ZTF-$g$ filter lightcurve of Chiron. \textit{(b):} ZTF-$g$ filter phase curve of Chiron. \textit{(c):} ZTF-$r$ filter lightcurve of Chiron. \textit{(d):} ZTF-$r$ filter phase curve of Chiron. All plots are color-coded for the same as Figure \ref{ChironATLASLightCurveCyanOrange}. $1\sigma$ uncertainty error bars are plotted though may be smaller than plot markers. \rcom{All observing seasons are labeled as per Table \ref{ChironObservingSeasons}.}
}
\label{ChironZTFLightCurveGreenRed}
\end{figure}

\begin{figure}
\centering
\includegraphics[width=\textwidth]{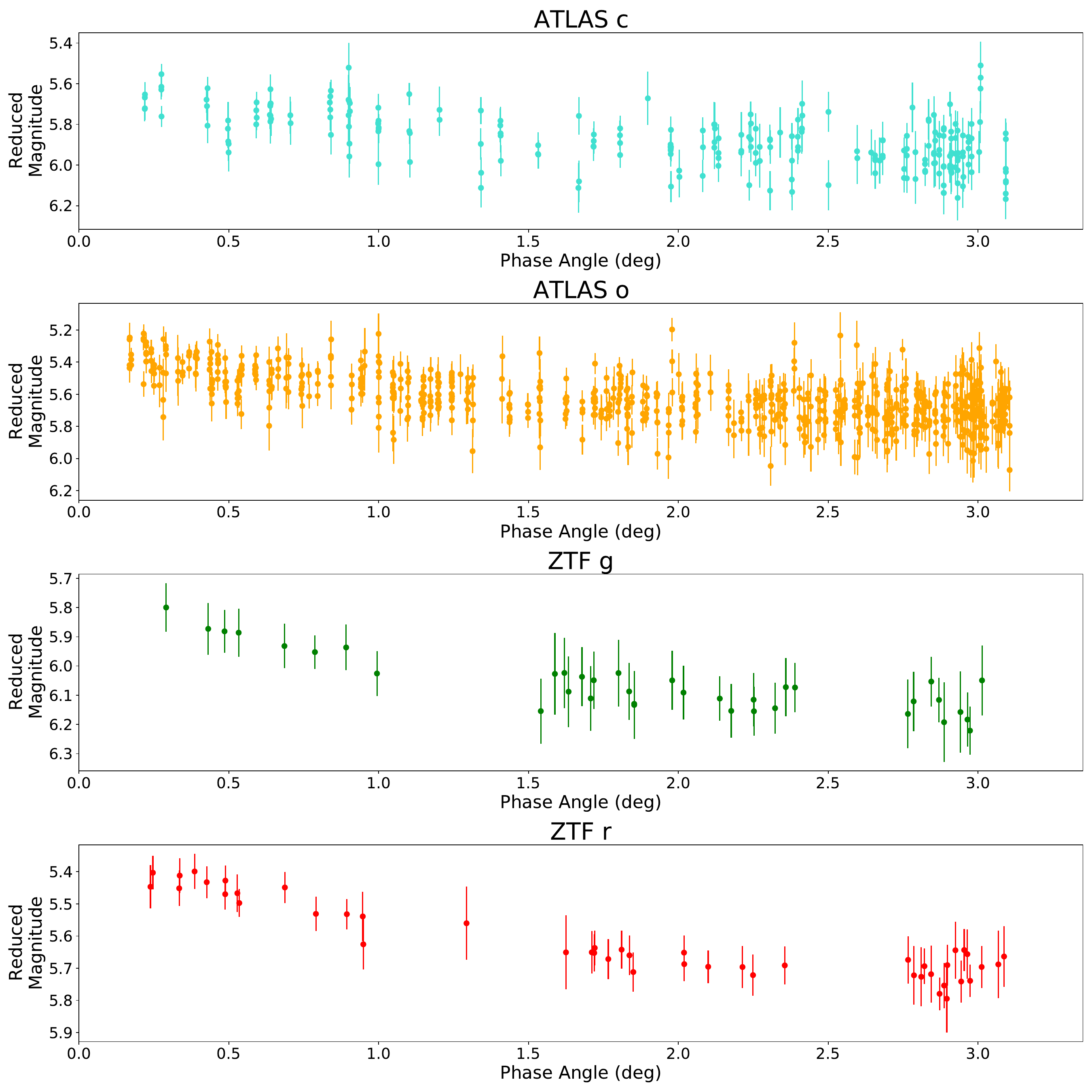}
\caption{Reduced magnitudes vs. solar phase angle of Chiron from ATLAS and ZTF from pre-2021 Brightening Event. Error bars are $1\sigma$ uncertainties.}
\label{ChironATLASZTFPhaseCurves}
\end{figure}

We first look at ATLAS and ZTF observations because they comprise the largest two datasets in our sample and cover the widest timespan. 
Figure \ref{ChironDataLightcurve} plots our ATLAS and ZTF reduced magnitudes vs. time\rcom{, both as measured in each filter and corrected for wavelength to the ATLAS $c$ and $o$ filters (due to them having the largest numbers of observations) using Chiron's pre-Brightening Event $c-g$ and $o-r$ colorindices calculated from the absolute magnitude values of Chiron in each filter.}
The reduced magnitudes separated by both survey and filter are shown in Figures \ref{ChironATLASLightCurveCyanOrange} (subplots \textit{a} and \textit{c}) and  \ref{ChironZTFLightCurveGreenRed} (subplots \textit{a} and \textit{c}). 
The `see-saw' motion of Chiron's brightness over time in each observing season is due to phase curve effects, with the brightness maxima corresponding to the smallest phase angles, and the spread of data is partly due to the periodic brightness variation caused by the Centaur's rotational lightcurve of amplitude ${\Delta}m\lesssim0.1$ mag; \citep[]{1989Icar...77..223B,2004A&A...413.1163G,2015A&A...576A..18O,2018MNRAS.475.2512C}.
As seen in Figure \ref{ChironDataLightcurve}, Chiron's brightness varies due to phase angle by approximately the same amount across observing seasons G-L (2015-2021), with peak reduced magnitude values in each filter consistent between observing seasons. \rcom{These reduced magnitude values are plotted against phase angle in Figure \ref{ChironATLASZTFPhaseCurves}. Fitting the phase curve of Chiron in each of these observing seasons separately, we find that Chiron's phase coefficient $\beta$ is consistent between observing seasons, implying Chiron's phase curve profile has been constant before the 2021 Brightening Event (see Section \ref{PhaseCoefficients} for more details).}
However, in observing season M (2021-2022) corresponding to the 2021 Brightening Event, Chiron suddenly increases in brightness by ${\sim}1$ mag compared to all previous observing seasons. The ATLAS and ZTF observations allow us to constrain the start date of the 2021 Brightening Event to between 2021 February 6 and 2021 June 17. Chiron passed aphelion between these dates, though observations were precluded due to the Centaur being behind the Sun as viewed from Earth.
This observed brightness increase far exceeds Chiron's brightness variation due to rotational modulation (${\lesssim}0.1$ mag) and solar phase angle (${\sim}0.5$ mag), meaning neither Chiron's lightcurve nor its phase curve can account for it. 
Furthermore, Chiron's brightness has continued to evolve. We see from Figure \ref{ChironDataLightcurve} that Chiron's peak magnitude dimmed significantly by ${\sim}0.5$ mag between observing seasons M (2021-2022) and N (2022-2023), but it continues to remain at brighter magnitudes than compared to G-L (2015-2021) by ${\sim0.5}$ mag.

The phenomenon that caused the 2021 Brightening Event changed Chiron's phase curve profile as well as its absolute magnitude. \rcom{The color-corrected phase curve of Chiron before, during, and after the 2021 Brightening Event is plotted in Figure \ref{ChironATLASZTFPhaseCurvesColorCorrected}. These magnitudes were color-corrected to the ATLAS $c$ and $o$ filters using Chiron's pre-Brightening Event $c-g$ and $o-r$ color indices calculated from the absolute magnitude values of linear fits to Chiron's phase curve in each filter.}
Figures \ref{ChironATLASLightCurveCyanOrange} and \ref{ChironZTFLightCurveGreenRed} show the brightness evolution (subplots $a$ and $c$) and phase curve (subplots $b$ and $d$) of Chiron in the ATLAS and ZTF datasets, respectively, color-coded by pre-, during, and post-2021 Brightening Event, with the latter two time ranges divided into pre- and post opposition. 
The ATLAS phase curves of individual post-2021 Brightening Event observing seasons are shown in Figure \ref{ChironATLASLightCurvePhaseCurvesObservingSeasonsCyanOrange} for clarity. 
Chiron's phase curves in observing seasons M (2021-2022) and N (2022-2023) are distinct both from each other and compared to the previous five observing seasons pre-2021 Brightening Event, which all lie on the same profile as shown in \rcom{Figures \ref{ChironATLASLightCurveCyanOrange}, \ref{ChironZTFLightCurveGreenRed}, \ref{ChironATLASZTFPhaseCurvesColorCorrected},  and \ref{ChironATLASLightCurvePhaseCurvesObservingSeasonsCyanOrange}} (see Section \ref{PhaseCoefficients} for more details). \rcom{Figure \ref{ChironATLASZTFPhaseCurvesColorCorrected} also shows that Chiron's phase curve profile is consistent between different filters across the same observing seasons.}
Figures \ref{ChironATLASLightCurveCyanOrange}, \ref{ChironZTFLightCurveGreenRed}, and \ref{ChironATLASLightCurvePhaseCurvesObservingSeasonsCyanOrange} also reveal that Chiron's brightness has evolved significantly within the timespan of a single observing season. Post-opposition magnitudes in observing season M (2021-2022) are brighter than those pre-opposition at the same phase angles.  
Measuring this difference for the ZTF-$r$ filter, we find the post-opposition magnitudes to be brighter than pre-opposition by ${\sim}2\sigma$ on average. The ZTF-$g$ filter data post-opposition is too sparse to allow an accurate measurement, but are consistent with pre-opposition within uncertainties. 
Pre- and post-opposition magnitude values in observing season N (2022-2023) are also consistent in brightness.
Chiron's evolving brightness and phase curve prevents us from fitting its phase curve during the 2021 Brightening Event or after, precluding removal of phase angle effects.


\rcom{
\begin{figure}
\centering
\includegraphics[width=\textwidth]{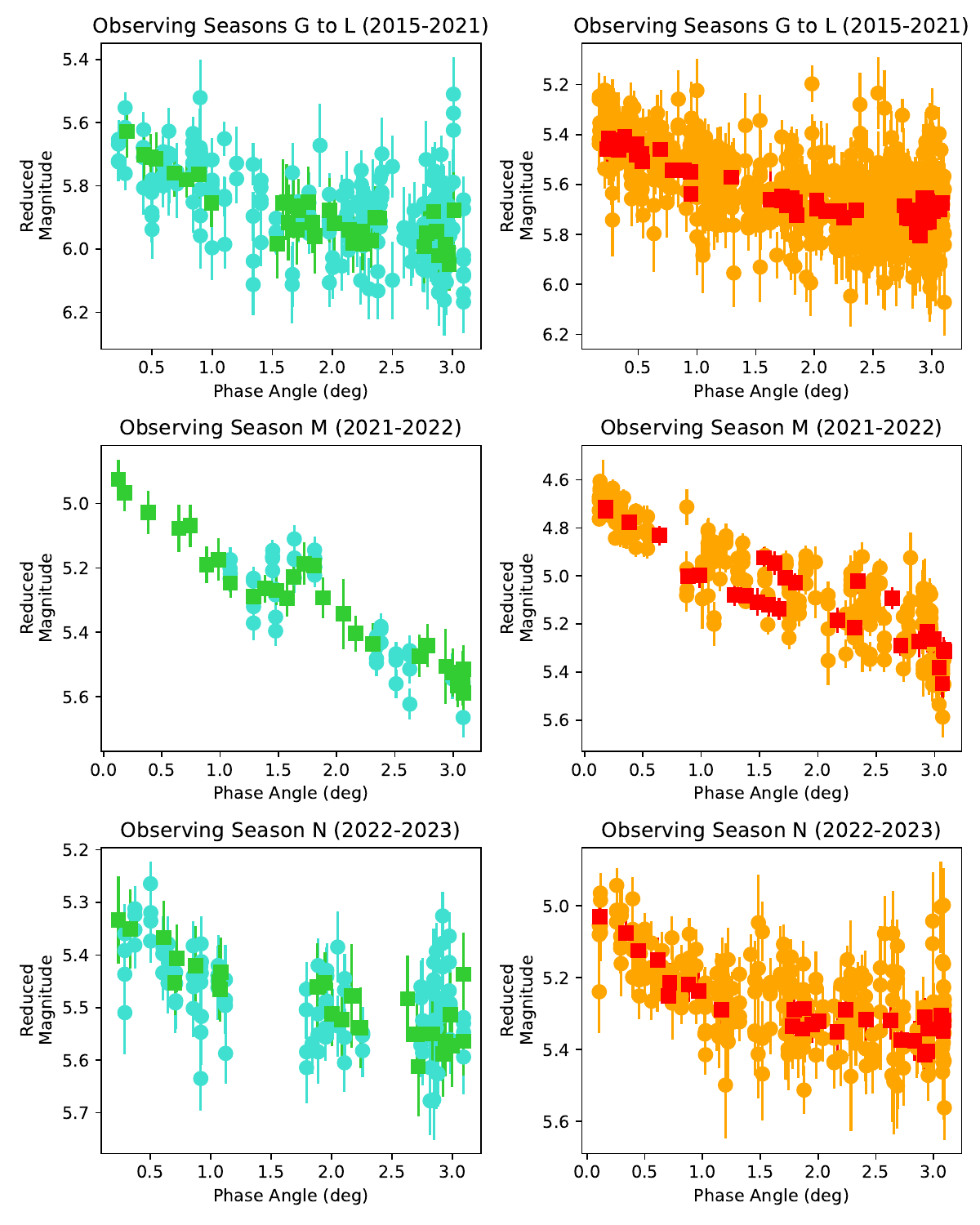}
\caption{\rcom{ATLAS c (turquoise circles) and ZTF g (green squares) color-corrected reduced magnitudes of Chiron vs. solar phase angle (left column) and ATLAS o (orange circles) and ZTF r (red squares) color-corrected reduced magnitudes of Chiron vs. solar phase angle (right column) for observing seasons G-L (2015-2021; top row), M (2021-2022; middle row), and N (2022-2023; bottom row). Error bars are $1\sigma$ uncertainties.}}
\label{ChironATLASZTFPhaseCurvesColorCorrected}
\end{figure}
}

\begin{figure}
\centering
\includegraphics[height=\textheight]{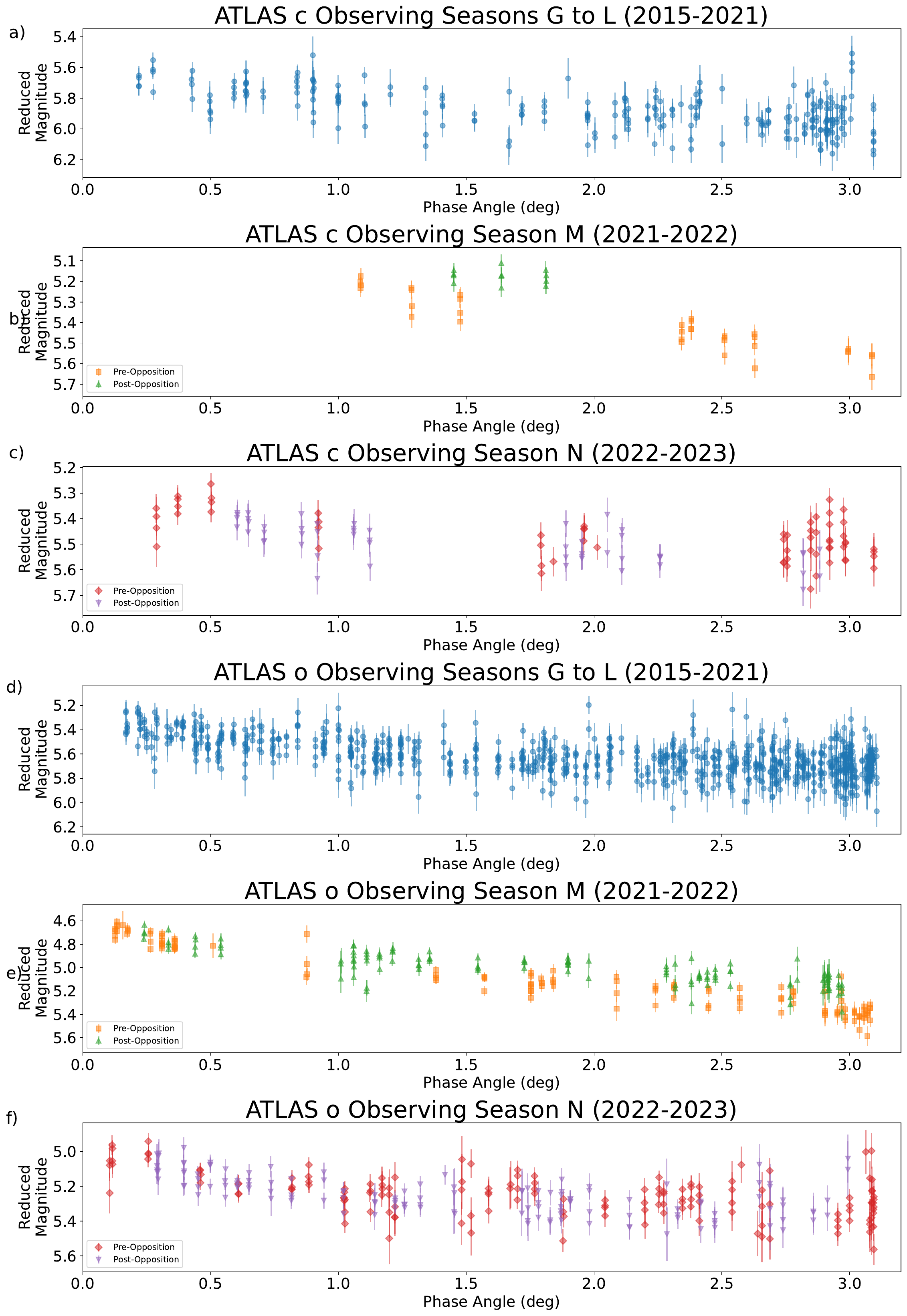}
\caption{ATLAS phase curves of Chiron across observing seasons in the $c$ filter (subplots \textit{a}, \textit{b}, and \textit{c}) and $o$ filter (subplots \textit{d}, \textit{e}, and \textit{f}), color-coded the same as Figure \ref{ChironATLASLightCurveCyanOrange}. Error bars are $1\sigma$ uncertainties.}
\label{ChironATLASLightCurvePhaseCurvesObservingSeasonsCyanOrange}
\end{figure}


\subsubsection{Phase Coefficients} \label{PhaseCoefficients}

The dense sampling coverage (see Figure \ref{ChironATLASZTFPhaseCurves}) of observing seasons I-L before the 2021 Brightening Event by ATLAS allows us to measure the linear phase coefficient of Chiron in each observing season utilizing the same methods and criteria of \citet{2023PSJ.....4...75D}. 
We find that Chiron's phase curve has remained relatively unchanged from 2012 up to the 2021 Brightening Event (observing seasons D-L). The $o$ filter linear phase coefficient values of each observing season I-M are listed in Table \ref{ChironBeta} and are consistent to $2\sigma$ within uncertainties, in accordance with the findings of \citet{2023MNRAS.523.3678B}. The comparatively sparser sampling of the ATLAS $c$ filter compared to $o$ means we can only accurately measure the linear phase coefficient for observing season L. Comparing the $c$ filter linear phase coefficient value for observing season L, $\beta_{c} = 0.110\pm0.018$ mag/deg, to that of the entire ATLAS baseline as measured by \citet{2023PSJ.....4...75D} ($\beta_{c} = 0.093\pm0.011$ mag/deg), we find them to be consistent within uncertainties. 
Figures \ref{ChironATLASLightCurveCyanOrange}, \ref{ChironZTFLightCurveGreenRed}, and \ref{ChironATLASLightCurvePhaseCurvesObservingSeasonsCyanOrange} highlight that Chiron's phase curve changed in shape during and post-2021 Brightening Event. This would cause different linear phase coefficient values when measured across pre-oppositon magnitude values compared to post-opposition ones and would lead to an inaccurate value when measured across the entire observing season. We therefore select only pre-opposition datapoints of observing season M (2021-2022) to reduce the effect of Chiron's evolving phase curve and fit a linear phase curve function to the ATLAS $o$ filter reduced magnitudes to quantify by how much the phase curve has changed. The pre-opposition ATLAS $c$ filter dataset for observing season M (2021-2022) is too sparse for an accurate phase curve fit. 
We find a linear phase coefficient value $\beta_{o} = 0.226 \pm  0.008$ mag/deg, differing significantly from the pre-2021 Brightening Event value measured by \citet{2023PSJ.....4...75D} ($\beta_{o} = 0.097\pm0.006$). \rcom{The presence of a dust coma may have caused this change in phase coefficient of Chiron, but the uncertainties and scatter of our ATLAS and ZTF observations prevent us from constraining the presence of a dust coma before Chiron's 2021 Brightening Event. Furthermore, although a standard phase curve model for cometary comae exists \citep{Schleicher_et_al_2010}\footnote{\url{https://asteroid.lowell.edu/comet/dustphase/details}}, it may not be applicable to Centaur comae if the mechanism which produces them or the type of dust emitted is different to comets. A detailed analysis of the effect of dust on Chiron's phase curve is therefore beyond the scope of this paper.}


\begin{deluxetable*}{lcr}[h] \label{ChironBeta}
\tablecaption{ATLAS $o$ filter phase coefficients for observing seasons pre- and post-2021 Brightening Event. Uncertainties are $2\sigma$ errors.}
\tablecolumns{3}
\tablehead{
\colhead{Observing Season} & 
\colhead{$\beta_{o}$} & 
\colhead{$N_{o}$} \\
\colhead{} & 
\colhead{(mag/deg)} &
\colhead{}
}
\startdata
I (2017-2018) &  $0.105 \pm  0.012$ &  235 \\
J (2018-2019) &  $0.115 \pm  0.013$ &  176 \\
K (2019-2020) &  $0.092 \pm  0.012$ &  178 \\
L (2020-2021) &  $0.093 \pm  0.011$ &  178 \\
M (pre-opposition 2021) &  $0.226 \pm  0.008$ &  108 \\
\enddata
\end{deluxetable*}


\subsection{Color Evolution} \label{ColourEvolution}


We use the dual filter coverage of ATLAS and ZTF datasets to look for any potential changes in Chiron's $c-o$ and $g-r$ colors across time. We cannot correct our reduced magnitudes for phase angle as Chiron's phase curve evolved significantly during observing seasons M (2021-2022, the Brightening Event) and N (2022-2023, post-Brightening Event). Therefore we utilize two methods to minimize phase angle effects on our observations. Firstly, for a given datapoint in one filter, we select the datapoint in the other filter closest in time, under the criterion that it is separated in time by ${\leq}2$ days (the approximate observation cadence of ATLAS) in order to identify any sudden changes in Chiron's color index. We use the reduced magnitude value of each observation to calculate Chiron's $c-o$ and $g-r$ color values and associated uncertainties. The second method involves fitting polynomial splines to its reduced magnitude values across time. 
By interpolating reduced magnitude measurements across time, we may better account for Chiron's evolving brightness and phase curve, including any sudden increase/decrease in magnitude. 
We generate $10^3$ synthetic lightcurves for each filter in a given observing season by offsetting each magnitude value by a random number generated from a Gaussian distribution with standard deviation equal to the magnitude uncertainty. We then fit 3rd-order polynomial splines to each synthetic lightcurve using the $interpolate$ package of the $scipy$ Python library \citep{2020NatMe..17..261V}. For a given datapoint of a given synthetic lightcurve in one filter, we extrapolate Chiron's brightness at that time in the other filter based on the fitted spline. We then use the resulting reduced magnitude values to calculate a $c-o$ and $g-r$ color value. We only perform this second method for observing seasons with ${\geq}25$ observations in a given filter, ensuring that the polynomial can fit Chiron's magnitude values accurately. 

\begin{figure}
\centering
\includegraphics[width=\textwidth]{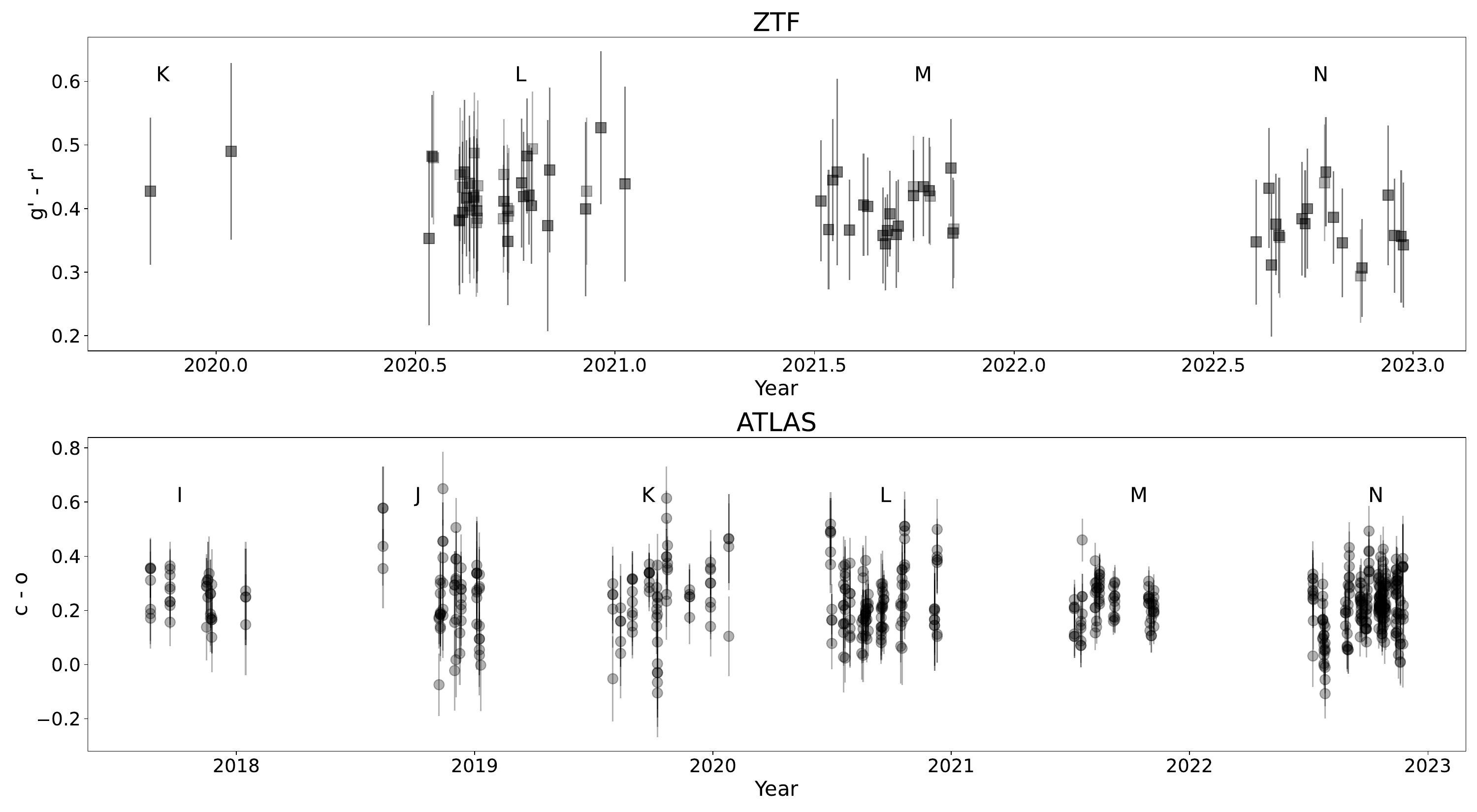}
\caption{ZTF (upper) and ATLAS (lower) color index values calculated from temporally adjacent data ($\leq2$ days apart) plotted across time. \rcom{All observing seasons are labeled as per Table \ref{ChironObservingSeasons}}. Error bars are $1\sigma$ uncertainties.}
\label{ChironColourEvolutionAdjacentDataATLASZTF}
\end{figure}

\begin{figure}
\centering
\includegraphics[width=\textwidth]{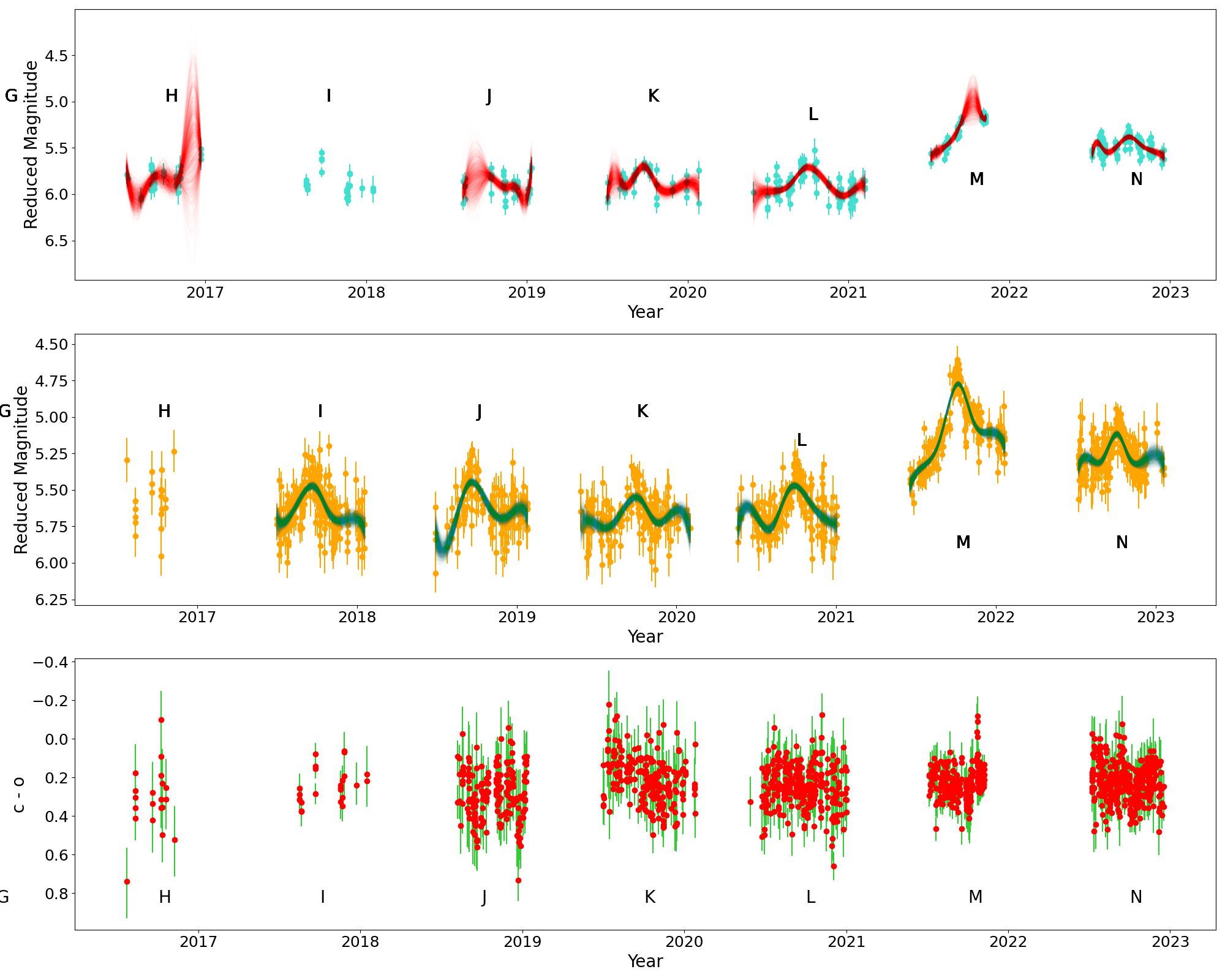}
\caption{
\textit{Upper plot:} ATLAS $c$ filter reduced magnitudes (turquoise circles) vs. time with 3rd order interpolated splines fitted to synthetic data overplotted (red). 
\textit{Centre plot:} ATLAS $o$ filter reduced magnitudes (orange circles) vs. time with 3rd interpolated splines fitted to synthetic data overplotted (teal). 
\textit{Lower plot:} Color index value calculated from the predicted magnitudes in each filter from the synthetic splines plotted across time. Red circles denote nominal color index values with green lines highlighting the range that includes the central 68.3\% of synthetic values, therefore corresponding to $1\sigma$ uncertainties.  \rcom{All observing seasons are labeled as per Table \ref{ChironObservingSeasons}}.}
\label{ChironColourEvolutionSplinesATLAS}
\end{figure}

\begin{figure}
\centering
\includegraphics[width=\textwidth]{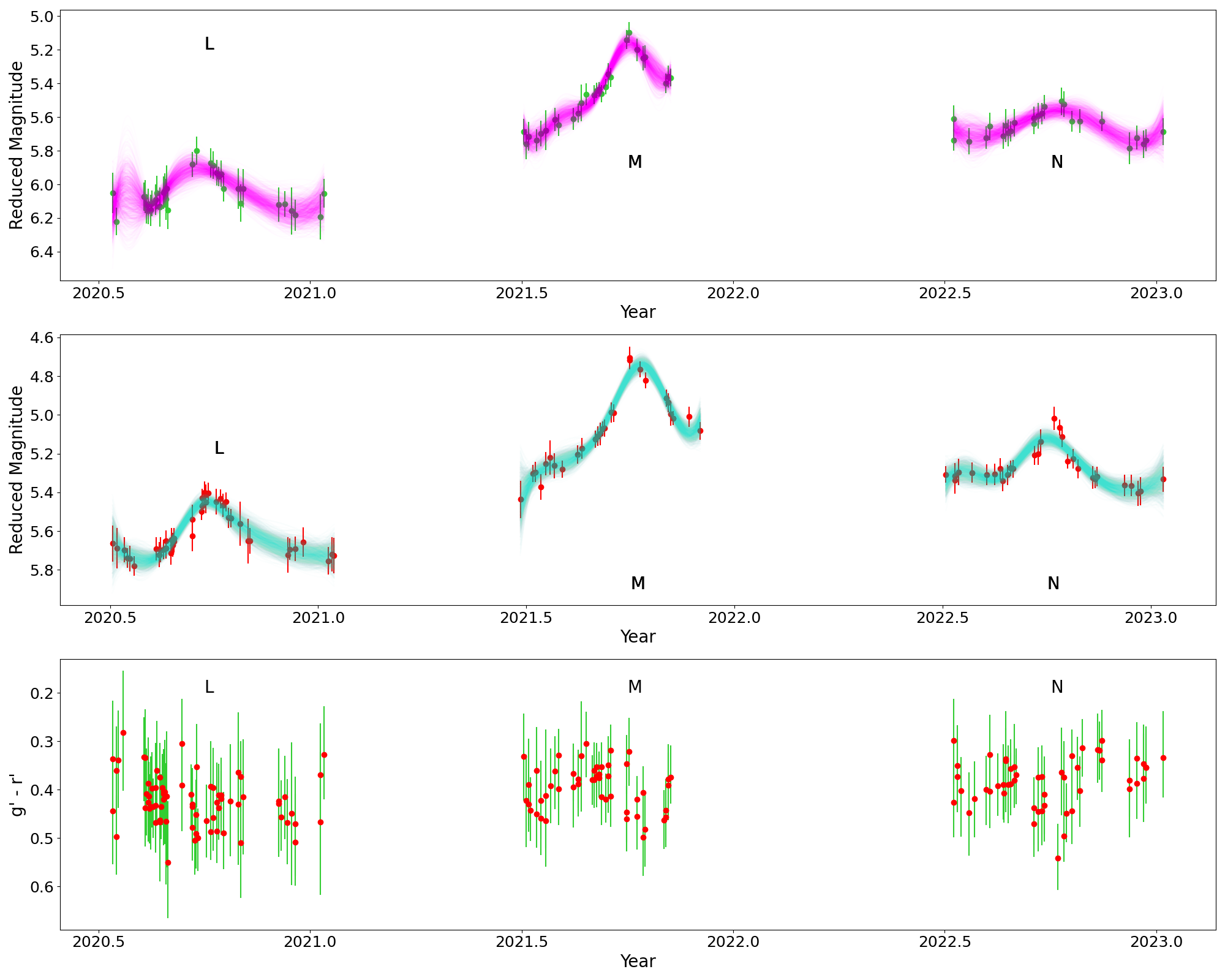}
\caption{
\textit{Upper plot:} ZTF-$g'$ filter reduced magnitudes (green circles) vs. time with interpolated 3rd order splines fitted to synthetic data overplotted (magenta). 
\textit{Centre plot:} ZTF-$r'$ filter reduced magnitudes (red circles) vs. time with interpolated 3rd order splines fitted to synthetic data overplotted (turquoise). 
\textit{Lower plot:} Color index value calculated from the predicted magnitudes in each filter from the synthetic splines plotted across time. Red circles denote nominal color index values with green lines highlighting the range that includes the central 68.3\% of synthetic values, therefore corresponding to $1\sigma$ uncertainties.  \rcom{All observing seasons are labeled as per Table \ref{ChironObservingSeasons}}.}
\label{ChironColourEvolutionSplinesZTF}
\end{figure}

Figure \ref{ChironColourEvolutionAdjacentDataATLASZTF} shows the resulting ATLAS $c-o$ and ZTF $g-r$ color indices of Chiron across time, calculated from temporally-adjacent points.
The ATLAS $c-o$ and ZTF $g-r$ values as calculated from the spline-fitting method are shown in Figures \ref{ChironColourEvolutionSplinesATLAS} and \ref{ChironColourEvolutionSplinesZTF} respectively. We find that Chiron has not undergone any significant color evolution across the baselines of ATLAS and ZTF, as Figures \ref{ChironColourEvolutionAdjacentDataATLASZTF}, \ref{ChironColourEvolutionSplinesATLAS}, and \ref{ChironColourEvolutionSplinesZTF} show that Chiron's color has remained constant within uncertainties across this time. 


Simultaneous $g$ and $r$ band measurements of Chiron were also obtained from our Gemini observations at three epochs (see Table \ref{tab:LLP-obs}). 
The sparse number of data points from Gemini combined with the evolution of Chiron's phase curve means we can only compare the Gemini $g-r$ values within a season. The two datapoints from season K are separated in phase angle by $\lesssim0.2$ deg, allowing comparison as brightness variation due to phase angle will be minimal. The distance-corrected $g-r$ values of the two datapoints of observing season N (2022-2023), $g-r = 0.501 \pm 0.057$ and $g-r = 0.557 \pm 0.052$, and are consistent to $1.0\sigma$.

\subsection{Search for Coma and PSF Extension} \label{ComaSearch}

\begin{figure}
\centering
\includegraphics[width=\textwidth]{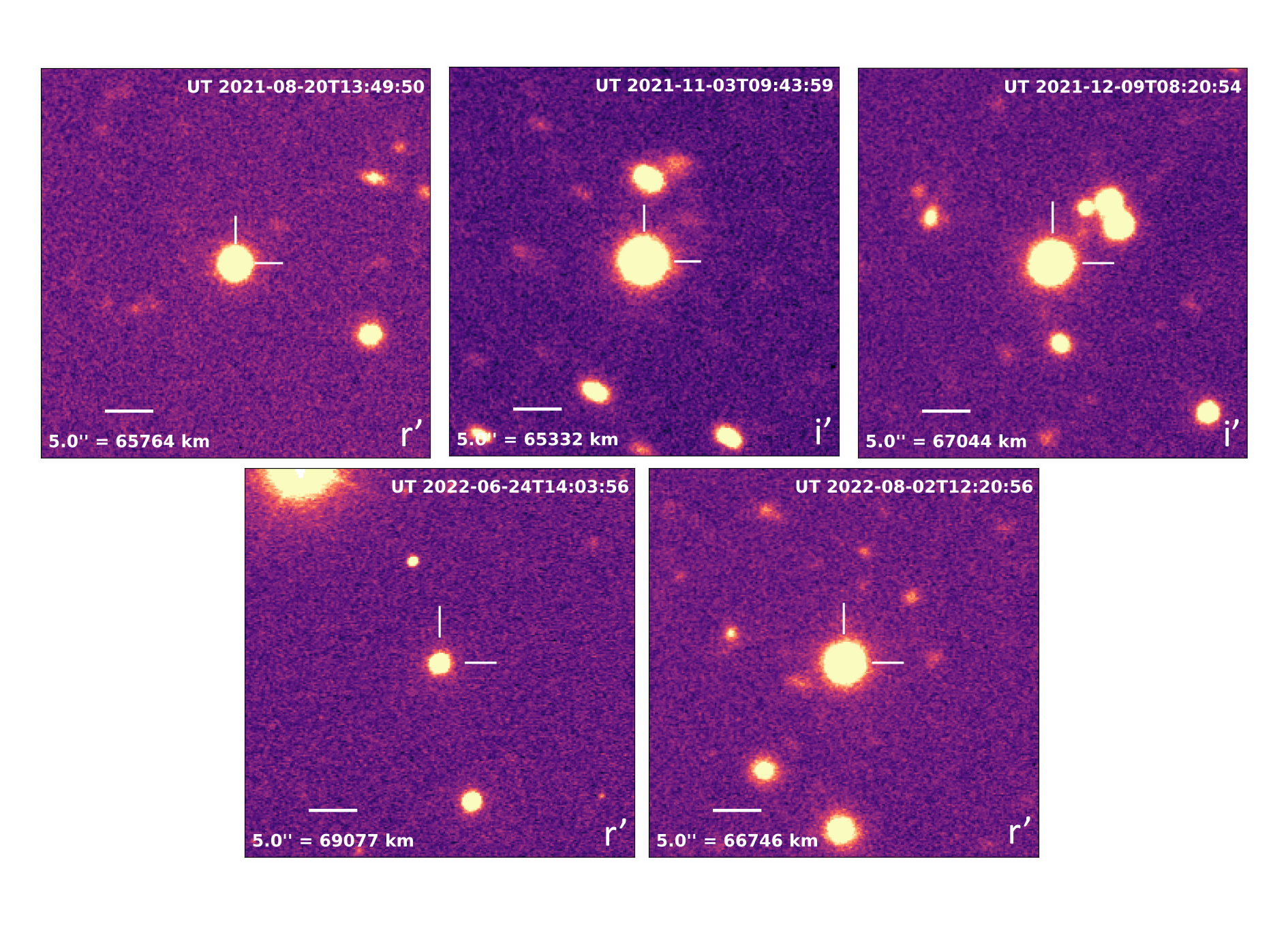}
\caption{Gemini stacked observations of Chiron in the $r'$ and $i'$ filters at all five epochs of observation. Cross-hairs indicate Chiron's position in each of the images. No coma or tails are visible in any of the images. The varying size of Chiron in each image is due to seeing variations and also relative image scales between individual nights. Note: Equatorial north is up and east to the left.}
\label{ChironGeminiObservation}
\end{figure}

\begin{figure}
\centering
\includegraphics[width=\textwidth]{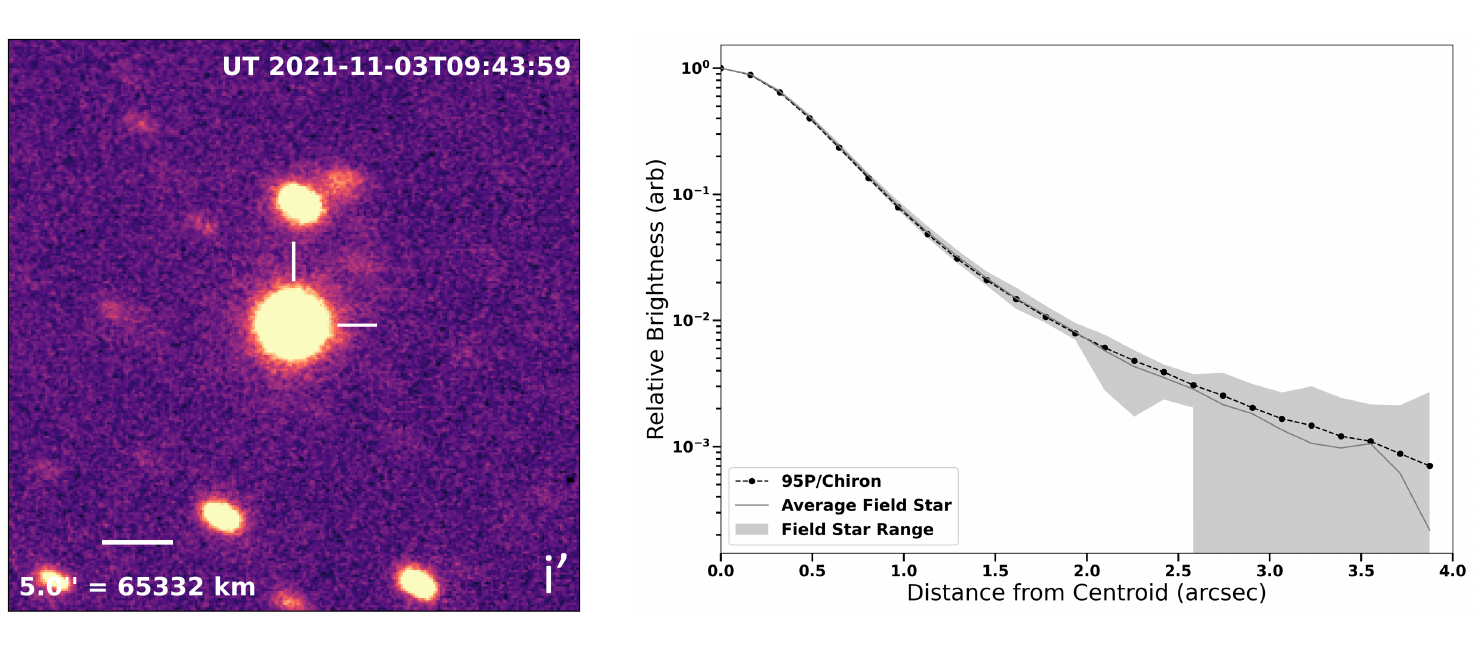}
\caption{Gemini stacked observation of Chiron in the $i'$ filter for 2021 November 3 with corresponding PSF radial profiles of Chiron and background field stars. Cross-hairs indicate Chiron's position in the image. PSF profile of Chiron is consistent with those of background stars.}
\label{ChironGeminiObservationPSF}
\end{figure}

Beyond photometry, we also analyze our Gemini and TRAPPIST-South observations for visible coma and PSF extension. Chiron has previously exhibited visible coma during instances of  brightening \citep[]{1990Icar...83....1H,2001PSS...49.1325S,2002Icar..160...44D}. However, \citet{2021RNAAS...5..211D} found no visible coma or PSF extension in ATLAS and LOOK observations during Chiron's 2021 Brightening Event. 
We stack our Gemini observations to search for any faint coma or PSF extension exhibited by Chiron during or post-2021 Brightening Event. The stacked images at the coordinates of Chiron from each of the five epochs listed in Table \ref{tab:LLP-obs} are shown in Figure \ref{ChironGeminiObservation} in the $r$ and $i$ filters. 
Our filter choice is motivated by the physical reality that the $r$ and $i$ filters are free from the brightest gas emission features in comets (Figure 1 in \citealt{2004come.book..425F}); the only detection of optical gas emission at Chiron previously reported was the CN (0-0) band at 3883\AA \ by \citet{1991Sci...251..774B} during its active phase in 1990.
No coma or tails are visible in any of the stacked Gemini images. We search for PSF extension by measuring the radial profiles of Chiron's PSF in each stacked images and comparing the result with field stars. We utilize Source Extractor Python \citep[]{1996A&AS..117..393B,2016JOSS....1...58B} to detect background sources in our observations. We select field stars for comparison to Chiron on the basis that 1) the peak flux count of their PSF does not exceed 50,000 electrons to ensure they are not saturated; 2) they are distant from neighbouring stars by at least the outer radius of any applied circular annulus for photometry; 3) they are brighter than 22nd magnitude, to ensure they are clearly visible in our images; 4) the uncertainties of both their catalog magnitude and the difference between their instrumental magnitudes and their Pan-STARRS1 catalog magnitudes are both less than 0.075 mag; 5) their catalog $g-r$ colors is within the range $-0.5 \leq g-r \leq 2.0$; 6) the difference between their catalog PSF magnitudes and Kron magnitudes in the $i$ filter is less than 0.05 mag so that we can separate stars from galaxies; and 7) their residuals to an unweighted linear fit to their difference in instrumental and catalog magnitudes as a function of catalog color index $g-i$ are less than $2.5\sigma$. We utilize the \textit{RadialProfile} function from the \textit{photutils} Python library \citep{larry_bradley_2023_7946442} to measure the radial profiles of Chiron and the background field stars. We apply a circular aperture of radius 13 pixels to both Chiron and the background field stars. To remove flux contribution from the background, we utilize a background annulus aperture of inner and outer radii 21.4 and 35.7 pixels, respectively, calculating the total background value from the product of the annulus area and the $3\sigma$-clipped median background value from the annulus. We subtract this total background flux value from the corresponding radial profile. An example of the resulting radial profiles is shown in Figure \ref{ChironGeminiObservationPSF}, where no visible coma is detected. Chiron's PSF is consistent with those of the background stars in all filters for all five epochs of observation. 

We also searched for signs of activity in our TRAPPIST-South observations (pre-Brightening Event, spanning 2012-2015), whose relatively large datasets and timespan allow a probe into past cometary behaviour. Neither visible coma nor tails are visible in any of our TRAPPIST-South observations. We searched for PSF extension by comparing the PSF profiles of Chiron to those of the field stars, using the same method as \citet{2021MNRAS.505..245D}. We created two stacks for each night, a first one on Chiron, compensating for its proper motion, and a second one on the background field stars.  
We then computed the distances of each pixel from the centroids of the PSFs of Chiron and the background stars, respectively. 
The radial profiles of Chiron 
were found to be identical to the field stars for every epoch.  


\subsection{Rings} \label{Rings}

To ascertain if Chiron's reported ring system together with a changing viewing angle as seen from Earth could explain the Centaur's observed brightness evolution we reconstruct the ring model as constructed by \citet{2015A&A...576A..18O}. We include flux contributions from Chiron's nucleus, its two rings as reported by \citet{2015A&A...576A..18O} based on observations of Chiron's 2011 occultation, and the 1970 cometary outburst, utilising a ring albedo\footnote{\citet{2015A&A...576A..18O} quotes the albedo value of Chiron's rings as $p_{V}=0.17$, however we can only reproduce their model using an albedo value of $p_{V}=0.18$.} value of $p_{V} = 0.18$ and ring widths of 6.6 km and 4.7 km (Private communication with [Anonymized for dual-anonymous review]). We calculate the aspect angle $\delta$ of Chiron at a given time and corresponding ecliptic coordinates according to the equation:
\begin{equation}
    \delta = \frac{\pi}{2} - \arcsin[\sin(\beta_{e})\sin(\beta_{p}) + \cos(\beta_{e})\cos(\beta_{p})\cos(\lambda_{e}-\lambda_{p})]
\end{equation}
where $\beta_{e}$ and $\lambda_{e}$ are the ecliptic latitude and longitude respectively of Chiron as observed from Earth and $\beta_{p}$ and $\lambda_{p}$ are the ecliptic latitude and longitude of the pole orientation of Chiron and its ring system. We then calculate the projected area $A_{p}(\delta)$ of Chiron at a given epoch and corresponding aspect angle $\delta$ according to the equation:
\begin{equation}
    A_{p}(\delta) = \frac{Area_{max}(\delta) + Area_{min}(\delta)}{2}
\end{equation}
where $Area_{max}(\delta)$ and $Area_{min}(\delta)$ are the maximum and minimum projected area defined by the following equations:
\begin{equation}
    Area_{max}(\delta) = \pi a[b^{2}\cos^{2}(\delta) + c^{2}\sin^{2}(\delta)]^{1/2}
\end{equation}
and 
\begin{equation}
    Area_{min}(\delta) = \pi b[a^{2}\cos^{2}(\delta) + c^{2}\sin^{2}(\delta)]^{1/2}
\end{equation}
where $a$, $b$, and $c$ are the semimajor axes of Chiron's nucleus assuming a triaxial ellipsoid shape in hydrostatic equilibrium. 
We use the same equation as \citet{2015A&A...576A..18O} to calculate the ratio of Chiron's total computed flux $F_{Chiron}$ to that of the Sun $F_{Sun}$ at a given time $t$ and corresponding aspect angle $\alpha$, given by:

\begin{equation}
    \frac{F_{Chiron}}{F_{Sun}} = A_{p}p_{V}f(\alpha) + p^{Ring1}_{V}f'(\alpha)\pi ((a_{1}+W_{1})^2 -a_{1}^2)\mu + p^{Ring2}_{V}f'(\alpha)\pi ((a_{2}+W_{2})^2 -a_{2}^2)\mu + A_{c}p_{d}exp(-(t-t_{0})/\tau_{d})
\end{equation}
where $A_{p}$ is the projected area of Chiron's nucleus; $p_{V}$ the geometric albedo of the nucleus; $f(\alpha)$ the solar phase function of the nucleus (taken to equal 1); $p^{Ring1}_{V}$ and $p^{Ring2}_{V}$ the geometric albedos of the first and second rings respectively; $f'(\alpha)$ the solar phase function of the rings (taken to equal 1); $\mu$ the absolute value of the cosine of the rings' aspect angle; $\pi ((a_{i}+W_{i})^2 -a_{i}^2)$ is the projected area of the $i$ ring, where $W_{i}$ is the width of the $i$ ring and $a_{i}$ the radial distance of each ring to Chiron's nucleus; $A_{c}$ is the initial scattering cross-section of the coma caused by Chiron's 1970 cometary outburst; $t_{0}$ is the initial time at which the coma dust cloud starts to decay; and $\tau_{d}$ is the decay time of the coma. We utilize the same parameter values of \citet{2015A&A...576A..18O} for the cometary outburst contribution. We then compute Chiron’s resulting absolute magnitude $H_{V}$ across time using the following equation:
\begin{equation}
    H_{V} = H_{V,Solar} + 2.5log_{10}(\pi\frac{(1.496\times10^{8})^{2}}{(F_{Chiron}/F_{Sun})})
\end{equation}
where $H_{V,Solar}$ is the absolute magnitude of the Sun, which we take to equal $-26.76$ \citep{2018ApJS..236...47W}.

We also collate all absolute magnitude measurements of Chiron from the literature \citep[]{2001Icar..150...94B,2001PSS...49.1325S,2002Icar..160...44D,2003A&A...400..369R,2003Icar..166..195B,2004A&A...413.1163G,2006A&A...450.1239B,2010Icar..210..472B,2013A&A...555A..15F,2016Ap&SS.361..212G,2018MNRAS.475.2512C}
applying a consistent phase angle correction to all values utilizing the $HG$ phase curve model with slope parameter $G = 0.70$ as determined by \citet{2001Icar..150...94B}. 
We transform the magnitude measurements from our own observations to the Johnson Cousins $V$-filter. For a given filter, we calculate the mean reduced magnitude of any observations taken before the 2021 Brightening Event. We then calculate the mean reduced magnitude of any $V$-filter observations (our own or from the literature) that coincide in time. The difference between these means is subtracted from every observation in the given filter. No $V$-filter measurements exist post-2021 Brightening Event, so we use the filter-corrected ATLAS $o$ observations, divided by pre- and post-opposition in each observing season, to transform the Gemini and LOOK magnitudes. We transform our filter-corrected reduced magnitudes to absolute magnitudes $H_{V}$ by applying a phase angle correction using $G = 0.70$ as per \citet{2001Icar..150...94B}. All the collated absolute magnitude values of Chiron, both from this work and from previous literature studies, are listed in Table \ref{ChironHV}.


\begin{deluxetable*}{lcccccccccr}[h] \label{ChironHV}
\tablecaption{Estimated Absolute Magnitudes of Chiron}
\tablecolumns{10}
\tablehead{
\colhead{UT Year}&
\colhead{Geocentric}&
\colhead{Heliocentric}&
\colhead{Phase}&
\colhead{Apparent}&
\colhead{Apparent}&
\colhead{Reduced}& 
\colhead{Filter}&
\colhead{Estimated}&
\colhead{Estimated}&
\colhead{Reference}\\
\colhead{}&
\colhead{distance}&
\colhead{distance}&
\colhead{angle}&
\colhead{Magnitude}&
\colhead{Magnitude}&
\colhead{Magnitude}&
\colhead{}&
\colhead{Absolute}&
\colhead{Absolute}&
\colhead{}\\
\colhead{}&
\colhead{}&
\colhead{}&
\colhead{}&
\colhead{}&
\colhead{Uncertainty}&
\colhead{}&
\colhead{}&
\colhead{Magnitude}&
\colhead{Magnitude}&
\colhead{}\\
\colhead{}&
\colhead{}&
\colhead{}&
\colhead{}&
\colhead{}&
\colhead{}&
\colhead{}&
\colhead{}&
\colhead{}&
\colhead{Uncertainty}&
\colhead{}\\
\colhead{}&
\colhead{(au)}&
\colhead{(au)}&
\colhead{(deg)}&
\colhead{}&
\colhead{}&
\colhead{}&
\colhead{}&
\colhead{}&
\colhead{}&
\colhead{}\\
}
\startdata
2023.098537 & 19.343 & 18.807 &2.485 & 18.111 &0.055 & 5.3068 & $r’$ & 5.2661 & 0.055 & This work \\
2023.098531 & 19.343 & 18.807 &2.485 & 18.475 &0.079 & 5.6708 & $g’$ & 5.2734 & 0.079 & This work \\
2023.098524 & 19.343 & 18.807 &2.486 & 18.127 &0.056 & 5.3228 & $r’$ & 5.2821 & 0.056 & This work \\
2023.098518 & 19.343 & 18.807 &2.486 & 18.485 &0.078 & 5.6808 & $g’$ & 5.2834 & 0.078 & This work \\
2023.093077 & 19.314 & 18.807 &2.540 & 18.125 &0.063 & 5.3240 & $r’$ & 5.2812 & 0.063 & This work \\
2023.093071 & 19.314 & 18.807 &2.540 & 18.453 &0.074 & 5.6520 & $g’$ & 5.2526 & 0.074 & This work \\
2023.093064 & 19.314 & 18.807 &2.540 & 18.164 &0.060 & 5.3630 & $r’$ & 5.3202 & 0.060 & This work \\
2023.093058 & 19.314 & 18.807 &2.541 & 18.556 &0.078 & 5.7550 & $g’$ & 5.3556 & 0.078 & This work \\
2023.076629 & 19.225 & 18.809 &2.689 & 18.071 &0.082 & 5.2799 & $r’$ & 5.2319 & 0.082 & This work \\
2023.076623 & 19.225 & 18.809 &2.689 & 18.611 &0.093 & 5.8199 & $g’$ & 5.4152 & 0.093 & This work \\
\enddata
\tablecomments{This table is published in its entirety in the machine-readable format. A portion is shown here for guidance for regarding its form and content.}
\tablecomments{The paper \citet{2001Icar..150...94B} does not provide values for apparent magnitude, its associated uncertainty, or reduced magnitude.}
\end{deluxetable*}

\begin{figure}
\centering
\includegraphics[width=\textwidth]{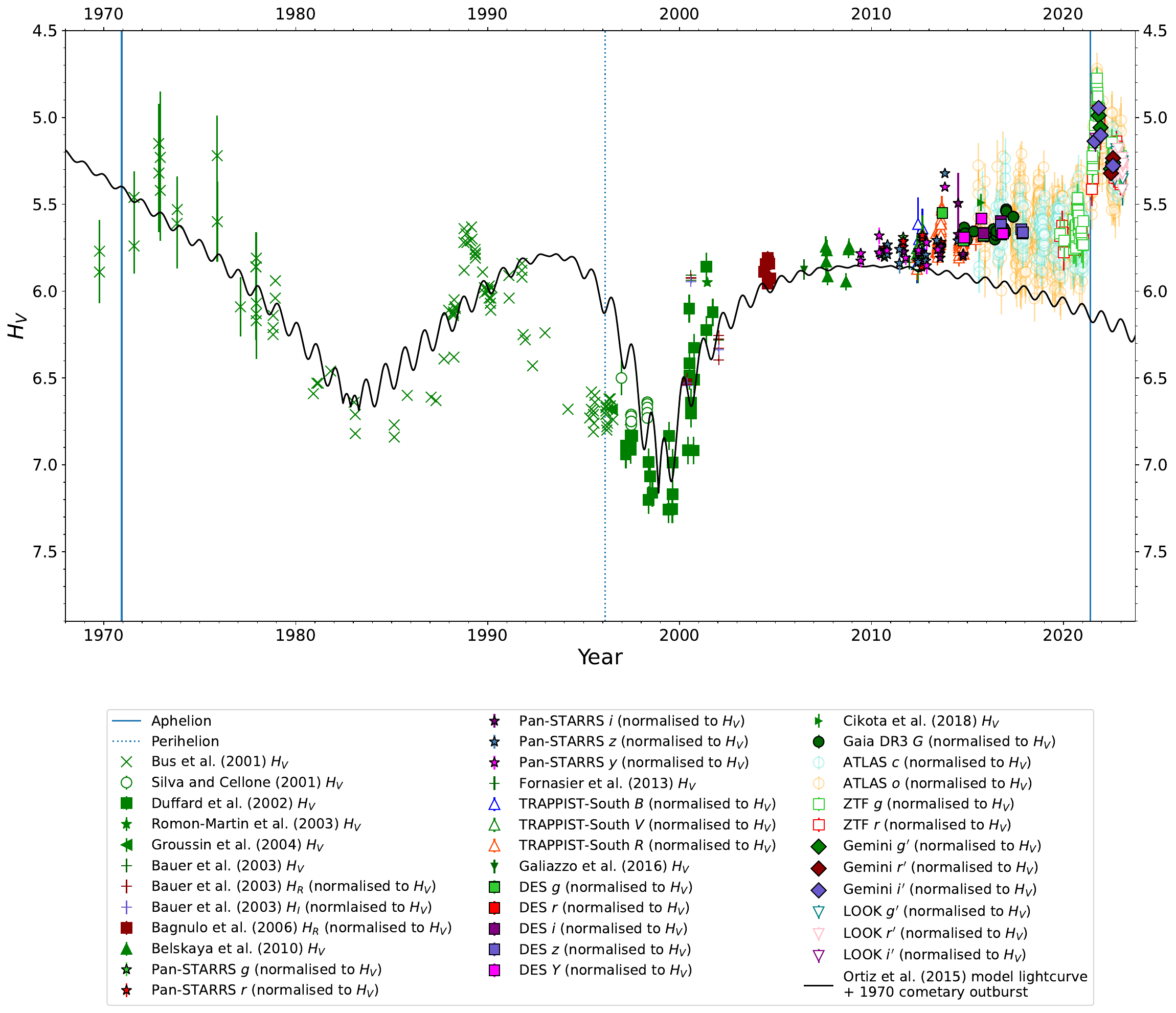}
\caption{Estimated $V$-band absolute magnitude (corrected for distance, phase angle, and filter color difference) vs. time of Chiron, with values obtained from our observations and literature measurements. Dates of aphelion and perihelion are indicated by the blue vertical solid and dotted lines, respectively. Chiron was known to exhibit visible coma during its outburst commencing around 1989 \citep[]{1989IAUC.4770....1M,1990Icar...83....1H,1990IAUC.4947....3M}, with later confirmed detections of coma in 1998 \citep{2001PSS...49.1325S}. \rcom{Chiron was observed to dim in brightness as it approached perihelion in 1996 February, reaching its dimmest recorded absolute magnitude shortly after perihelion around 1999.} Other likely cometary outbursts are thought to have occurred in the 1970s \citep{2001Icar..150...94B} and around the year 2000 \citep{2002Icar..160...44D} as evidenced by Chiron's sudden increases in absolute magnitude around these times. The lightcurve as predicted from the ring model of \citet{2015A&A...576A..18O} is plotted in black. Chiron's 2021 Brightening Event contrasts with the gradual dimming as predicted by \citet{2015A&A...576A..18O}.}
\label{ChironRingModelLightCurve}
\end{figure}

\begin{figure}
\centering
\includegraphics[width=\textwidth]{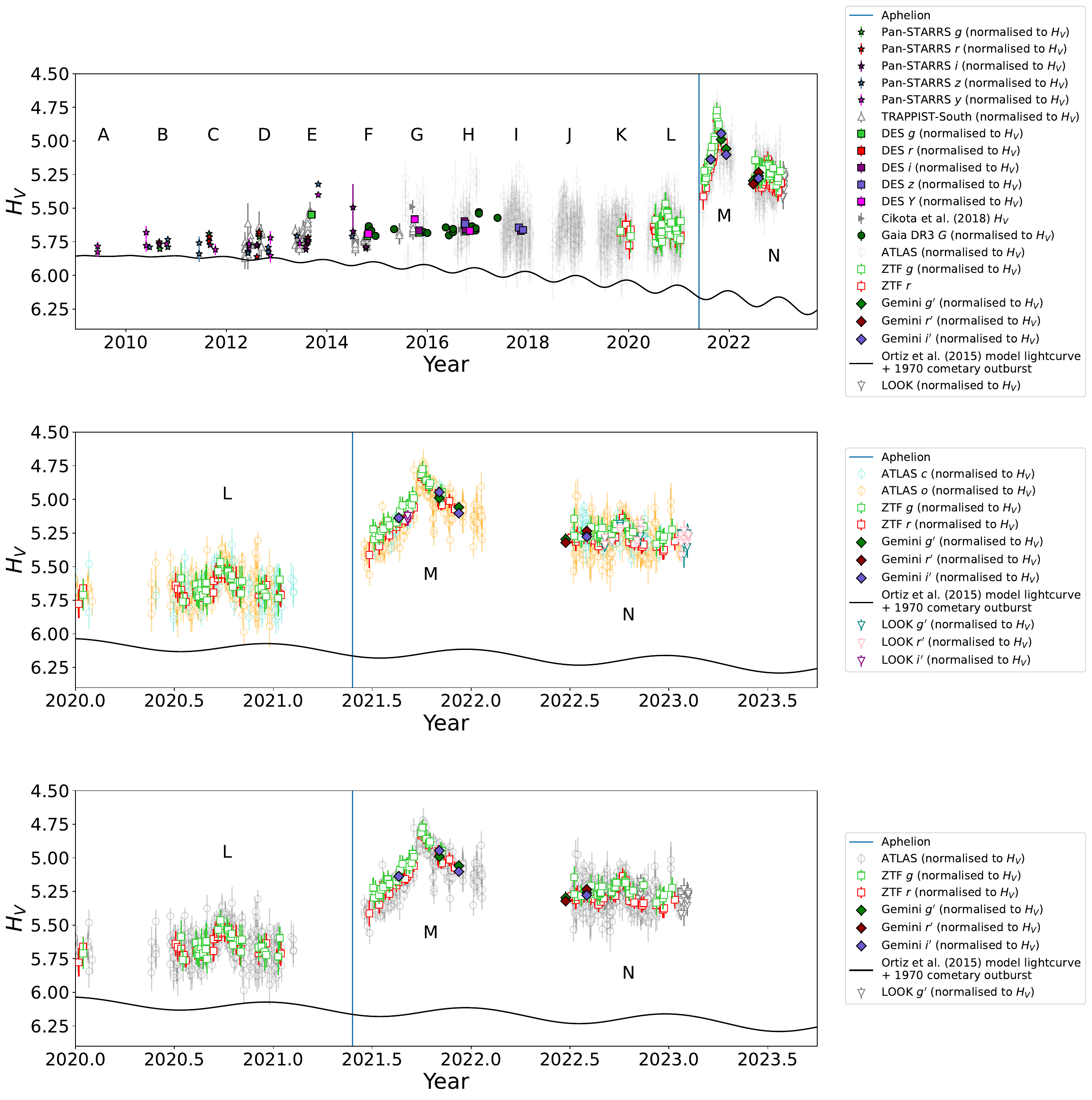}
\caption{\textit{Upper plot:} Estimated $V$-band absolute magnitude (corrected for distance, phase angle, and filter color difference) vs. time for Chiron across our baseline of observations, with the high-precision observations from Gemini, Gaia, DES, and ZTF highlighted for clarity. \textit{Centre plot:} Estimated $V$-band absolute magnitude (corrected for distance, phase angle, and filter color difference) vs. time for Chiron across observing seasons L-N (2020-2023) highlighting the brightening in Chiron's magnitude due to the 2021 Brightening Event. \textit{Lower plot:} Estimated $V$-band absolute magnitude (corrected for distance, phase angle, and filter color difference) vs. time for Chiron across observing seasons L-N (2020-2023) highlighting the brightening in Chiron's magnitude due to the 2021 Brightening Event. The observations from Gemini and ZTF are highlighted for clarity. \rcom{All observing seasons are labeled as per Table \ref{ChironObservingSeasons}.} In all subplots, the date of aphelion is indicated by the solid blue line. Chiron's absolute magnitude as predicted by \citet{2015A&A...576A..18O} is plotted in black.}
\label{ChironRingModelLightCurveOurObservations}
\end{figure}

Figure \ref{ChironRingModelLightCurve} shows the estimated absolute magnitude of Chiron over time combining our datasets and measurements from the literature as compared to \citet{2015A&A...576A..18O}. Chiron's absolute magnitude across our baseline of observations is shown in Figure \ref{ChironRingModelLightCurveOurObservations} for clarity.
As seen in Figures \ref{ChironRingModelLightCurve} and \ref{ChironRingModelLightCurveOurObservations}, the latter showing the post-Brightening Event era, Chiron's brightness evolution across our baseline of observations is inconsistent with that predicted by \citet{2015A&A...576A..18O}. The high-precision Pan-STARRS, DES, ZTF, and \textit{Gaia} measurements show that even before 2021, Chiron was already significantly brighter than predicted by \citet{2015A&A...576A..18O}. The ring model predicted Chiron to slowly dim across observing seasons M (2021-2022, the 2021 Brightening Event) and N (2022-2023, post-Brightening Event). However, Chiron's measured absolute magnitude from ATLAS, ZTF, LOOK, and Gemini was sigificantly brighter than the \citet{2015A&A...576A..18O} ring model during this time, differing by ${\sim}0.5-1$ mag and reaching a peak magnitude comparable to its historical maximum brightness in the 1970s \citep{2001Icar..150...94B}. The \citet{2015A&A...576A..18O} ring model fails to match Chiron's brightness evolution across our baseline of observations and therefore cannot explain the 2021 Brightening Event.

\section{Discussion} \label{Discussion}

\rcom{Chiron's observed increase in brightness from 2021 onwards far exceeds its amplitude of rotational modulation, predicted to be ${\lesssim}0.05$ mag during 2021 \citep{2015A&A...576A..18O} when phase-folded to its measured rotation period of $5.917813$ hr \citep[]{1989Icar...77..223B,1993Icar..104..234M}, and was too small to be measured in ATLAS data from \citet{2021RNAAS...5..211D}. Furthermore, Chiron's observed increase in brightness exceeds its brightness variation due to its phase function. We therefore rule out both Chiron's rotation and phase curve as \rcom{being} responsible for the 2021 Brightening Event.}
\citet{2023MNRAS.523.3678B} hypothesize that Chiron's observed evolution in brightness could instead be due to variation in surface albedo with areas of differing albedo becoming visible to Earth as the Centaur's aspect angle changes throughout its orbit. However, comparing pre- and post-opposition ATLAS and ZTF observations from observing season M (2021-2022), we have seen that Chiron's brightness evolved significantly, at an appreciable rate across the timespan of a single observing season. The short timescale of Chiron's observing season M (0.6 years) compared to its orbit around the Sun (50.8 years) means that Chiron has only swept through $1.2\%$ of its orbit during this time. \rcom{Combined with the measured ecliptic coordinates of Chiron's axial pole of $\lambda = 151 \pm 8$ deg and $\beta = 18 \pm 11$ deg \citep{2023arXiv230803458O} implying a near face-on obliquity to Earth, as well as the Centaur's small amplitude of rotational modulation, we consider it implausible for a surface feature on Chiron as suggested by \citet{2023MNRAS.523.3678B} 
to enter and exit view from Earth across such a small percentage of the object's orbit. We thus reject surface albedo variations as being responsible for Chiron's brightening and dimming across 2021-2023. }

We also consider the Chiron ring system model proposed by \citet{2015A&A...576A..18O} to be an unlikely cause of the object's observed brightening during and since 2021. Reconstructing this model, we find Chiron's measured absolute magnitudes to be significantly brighter than predicted by ${\sim}0.5-1$ mag. \rcom{Furthermore, the timescale on which Chiron's 2021 Brightening Event occurred is too short to be explained by the ring model, which predicts variation in Chiron's absolute magnitude across decadal timespans.} The \citet{2015A&A...576A..18O} lightcurve also fails to replicate the long-term evolution of Chiron's absolute magnitude both pre- and post-2021 Brightening Event. Chiron was predicted to slowly dim over time when it instead was observed to brighten in 2021. \rcom{Additionally}, Chiron was also significantly brighter pre-2021 Brightening Event than predicted by the ring model. 
We note, however, that although the \citet{2015A&A...576A..18O} model cannot explain \rcom{either the 2021 Brightening Event or Chiron's brightness evolution before this across our observation baseline,} this does not completely rule out the possibility that Chiron's rings are not well-described by current models and are at least partially responsible for this brightening in reduced magnitude. 
Analyses of stellar occultations by Chiron in 2018 and 2022 indicate evolution of the material surrounding Chiron on relatively short timescales \citep[]{2023A&A...676A..72B,2023arXiv230803458O,2023arXiv231016205S}.
\citet{2023arXiv231016205S} found Chiron's rings, as measured during the 2018 occultation, to be of reduced optical depth compared to 2011, with a reported feature, potentially indicative of a third ring, detected during immersion but not during emersion.
Furthermore, \citet{2023arXiv230803458O} detected signals indicative of additional, previously-undiscovered ring-like features during Chiron's 2022 occultation with the signals of the already-known rings being more pronounced compared to 2011, indicating a significant change in the ring system's configuration. Occultation studies also reveal that Chiron's rings are not azimuthally homogeneous in width \citep{2023arXiv230803458O} exhibiting agglomerations of ring particles possibly caused by small, unseen shepherd satellites. Collisions of satellites have been proposed as one mechanism for creating Centaur rings \citep{2014Natur.508...72B,2015A&A...576A..18O,2017A&A...602A..27M}, and such an event could act to augment Chiron's ring system, and thus its apparent brightness as seen from Earth. Future occultation studies are necessary to reveal the true nature of the reported ring system surrounding Chiron in order to better estimate their contribution to the Centaur's changing brightness across time.

Despite the need for an improved understanding of Chiron's rings, we argue that Chiron's observed evolution in brightness from 2021 onwards is \rcom{better described by} an epoch of cometary activity. The increase in brightness in 2021 followed by a dimming across time is consistent with Chiron's observed brightness evolution during previous cometary episodes (e.g. \citealt{1989Icar...77..223B,1989IAUC.4770....1M}).
This years-spanning change in brightness, though far longer than the weeks- \citep[]{2008A&A...485..599T,2016Icar..272..387M,2017Icar..284..359S,2019AJ....158..259S,2020tnss.book..307P} or months-long \citep[]{2006CBET..563....1C,2006IAUC.8656....2C,2011IAUC.9213....2J,2016MNRAS.462S.432R,2018JBAA..128...51J,2019AJ....157...88S,2019AJ....158..255K} cometary outbursts of the active Centaurs 29P/Schwassman-Wachmann 1 and 174P/Echeclus respectively, is nevertheless in keeping with the duration of previous epochs of Chiron's cometary activity, which have also lasted for years
\citep[]{1989Icar...77..223B,1989IAUC.4770....1M,1990Icar...83....1H,1990AJ....100..913L,1990IAUC.4947....3M,1990IAUC.4970....1W,1990AJ....100.1323M,1990nba..meet...83D,1988IAUC.4554....2T,2001Icar..150...94B,2001PSS...49.1325S}. 
Additionally, \citet{2023arXiv230803458O} theorized that material ejected due to an increase in cometary activity in 2021 could augment Chiron's rings, thereby explaining the apparent change in configuration in Chiron's ring system between occultations. Whether such a cometary outburst would represent an increase from quiescence or an entirely new epoch of cometary activity from a previously quiescent state is unclear. We did not detect any visible indicators of activity in the TRAPPIST-South observations spanning the years 2012-2015, yet \citet{2018MNRAS.475.2512C} detected a tentative ${\sim}5$ arcsec-length tail-like feature of surface brightness $25.3$ mag/arcsec${^2}$ in their 2015 observations of Chiron. \citet{2018MNRAS.475.2512C} also reported that Chiron exhibited fluctuations in brightness exceeding that of background field stars, implying low-level `microactivity'. Therefore, we consider the most likely cause of Chiron's 2021 Brightening Event to be an epoch of either new or increased cometary activity.

If Chiron is presently undergoing cometary activity, an explanation for why we do not detect any visible coma could be that it is too faint to be directly observed beyond the central point spread at Chiron's present heliocentric distance. Previous measurements of Chiron's coma report $V$-band surface brightness values of 24.6 mag/arcsec$^{2}$ \citep{1990IAUC.4970....1W} in 1990 when Chiron was 11.2 au from the Sun, and 26.0 mag/arcsec$^{2}$ \citep{2001PSS...49.1325S} in 1998 at a heliocentric distance of 8.938 au. 
We calculate the limiting magnitudes of our Gemini deep-stacked observations.
We utilize the same method of measuring the radial profiles of Chiron and the field stars as per Section \ref{ComaSearch}, and calculate the limiting surface brightness of each observation at a radial distance from the centroid of Chiron's PSF of three times the full-width half-maximum of a Gaussian fit to the PSF.
We re-calculate Chiron's literature coma surface brightness values to the corresponding heliocentric and geocentric distances for each Gemini observation. 
We transform the limiting surface brightnesses and Chiron's literature coma surface brightnesses $M_{surface}$ into total magnitudes $M$ according to the equation:

\rcom{
\begin{equation}
    M = M_{surface} - 2.5\log_{10}(2 \pi r^{2})
\end{equation}
}
\rcom{where $r$ is the radial distance of the surface brightness measurement from the PSF centroid \citep{2005MNRAS.358..641L}.}
\rcom{We then transform Chiron's coma values to the corresponding Pan-STARRS filter of each Gemini observation, transforming firstly from Johnson-Cousins to SDSS photometric systems and then from SDSS to Pan-STARRS. To transform from Johnson-Cousins to SDSS we utilize the transformation equations of \citet{2005AJ....130..873J} for stars with Johnson-Cousins color indices $R_{c}-I_{c}<1.15$ and $U-B\geq0$. These require $B-V$ and $R_{c}-I_{c}$ color index measurements of Chiron's coma; we use the $B-V=0.3$  measurement from \citet{1991A&A...241..635W} for this transformation, and in the absence of an historical $R_{c}-I_{c}$ measurement of Chiron's coma, we utilize the color index value $R-I=0.36$ \citep{2016Ap&SS.361..212G} of Chiron itself as a proxy. We then use the transformation equations of \citet{2016ApJ...822...66F} to transform these SDSS values into the Pan-STARRS photometric filter system. We utilize Chiron's $g-i$ color as measured in each epoch of Gemini observations; if an epoch lacks such a measurement, we instead use the closest $g-i$ measurement in time to that epoch.}
Table \ref{ChironGeminiLimitingMagnitudes} shows our resulting $3\sigma$ limiting magnitudes for each observation with Chiron's corresponding predicted coma brightness.
For all our Gemini observations, for both surface brightness and total magnitude, we find Chiron’s coma brightness to be fainter than the $3\sigma$ limiting magnitude of the observation, assuming a similar coma to that observed in 1990.

\begin{deluxetable*}{cccccccc} \label{ChironGeminiLimitingMagnitudes}
\tablecaption{$3\sigma$ Limiting Magnitudes of the Gemini Observations as Compared to the Predicted Brightness of Chiron's Coma.}
\tablecolumns{7}
\tablehead{
\colhead{Epoch} & 
\colhead{Radial} & 
\colhead{$3\sigma$ Limiting} & 
\colhead{$3\sigma$ Limiting} & 
\colhead{Coma Surface} & 
\colhead{Coma} & 
\colhead{Coma Surface} & 
\colhead{Coma} \\
\colhead{and} & 
\colhead{Distance} & 
\colhead{Surface} & 
\colhead{Magnitude} & 
\colhead{Brightness} & 
\colhead{Magnitude} & 
\colhead{Brightness} & 
\colhead{Magnitude} \\
\colhead{Filter} & 
\colhead{Distance} & 
\colhead{Brightness} & 
\colhead{} & 
\colhead{(W90)} & 
\colhead{(W90)} & 
\colhead{(SC01)} & 
\colhead{(SC01)} \\
\colhead{} & 
\colhead{\rcom{(arcsec)}} & 
\colhead{(mag/arcsec$^{2}$)} & 
\colhead{(mag)} & 
\colhead{(mag/arcsec$^{2}$)} & 
\colhead{(mag)} & 
\colhead{(mag/arcsec$^{2}$)} & 
\colhead{(mag)}
}
\startdata
2021-08-20 g & 2.8162 & 22.453 &18.209 & 25.706 &20.831 &27.796 & 21.317 \\
2021-08-20 r & 2.6550 & 24.449 &20.334 & 25.686 &20.810 &27.775 & 21.297 \\
2021-08-20 i & 2.5649 & 22.650 &18.609 & 25.371 &20.496 &27.461 & 20.982 \\
2021-11-03 g & 3.6173 & 24.840 &20.053 & 25.704 &20.814 &27.794 & 21.301 \\
2021-11-03 i & 2.8325 & 23.815 &19.559 & 25.371 &20.481 &27.461 & 20.968 \\
2021-12-09 g & 3.7897 & 23.756 &18.867 & 25.714 &20.880 &27.804 & 21.367 \\
2021-12-09 i & 3.2656 & 23.859 &19.294 & 25.371 &20.537 &27.461 & 21.024 \\
2022-06-24 g & 1.6347 & 24.037 &20.975 & 25.684 &20.915 &27.774 & 21.402 \\
2022-06-24 r & 1.4345 & 23.442 &20.663 & 25.681 &20.912 &27.771 & 21.399 \\
2022-08-02 g & 2.9405 & 24.923 &20.586 & 25.683 &20.840 &27.773 & 21.327 \\
2022-08-02 r & 2.6914 & 24.373 &20.228 & 25.681 &20.837 &27.770 & 21.324 \\
2022-08-02 i & 2.4350 & 23.670 &19.742 & 25.367 &20.524 &27.457 & 21.011 \\
\enddata
\tablecomments{Coma brightness values have been converted to the corresponding filter of each Gemini observation.}
\tablecomments{W90: \citet{1990IAUC.4970....1W}; SC01: \citet{2001PSS...49.1325S}}
\end{deluxetable*}

\begin{figure}
\centering
\includegraphics[width=\textwidth]{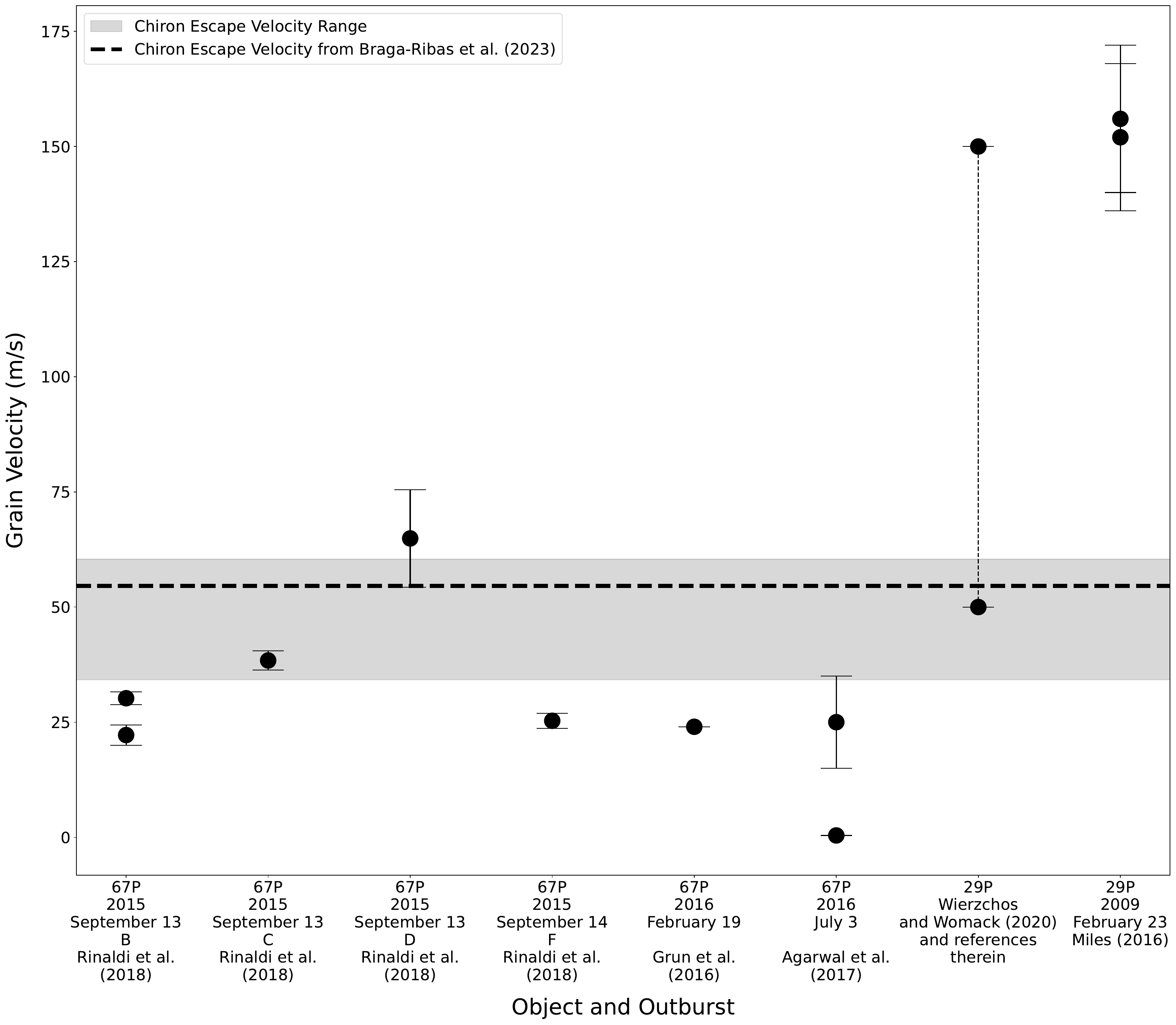}
\caption{Dust grain velocity measurements of outbursts from the Jupiter-family comet 67P/Churuymov-Gerasimenko compared to the range of escape velocity values of Chiron (grey) estimated from known densities \rcom{of} Centaur binaries. Horizontal dashed black line indicates value of Chiron's escape velocity according to the density estimate of \citet{2023A&A...676A..72B}.}
\label{ChironGrainVelocities}
\end{figure}

Alternatively, the lack of detected coma could be due to the mass of the Centaur's nucleus. The large size of Chiron (volume-equivalent radius ${\sim}98$km; \citealt{2023A&A...676A..72B}) compared to most cometary nuclei means its gravitational force is non-negligible. Some material emitted during an active epoch may be bound to the nucleus if its escape velocity exceeds the grain velocities of the emitted particles, preventing the formation of a spatially-extended coma detectable from Earth-based observations. A bound coma around Chiron was previously discovered in 1993 by \citet{1997AJ....113..844M} using the Hubble Space Telescope. These revealed that Chiron's brightness profile could only be explained by a coma \rcom{whose} structure comprised two components, one spatially extended and the other bound to the nucleus within its exopause \citep{1997AJ....113..844M}. To ascertain the possibility that an entirely bound coma could exist around Chiron, we estimate its nuclear escape velocity via the equation:

\begin{equation}
    v_{esc} = \sqrt{\frac{4\pi G\rho abc}{3r}}
\end{equation}
where $G$ is the gravitational constant; $\rho$ is the density of Chiron's nucleus; $a$, $b$, and $c$ are the semimajor axes of the nucleus assuming the triaxial ellipsoidal shape of \citet{2023A&A...676A..72B}; with $r$ being the equivalent volume radius of the nucleus. \citet{2023A&A...676A..72B} estimated Chiron's density to be $1.119$ g cm$^{-3}$ based on a Jacobi ellipsoidal model of its shape derived from stellar occultations under the assumption of hydrostatic equilibrium. As this value is model-dependent, we also estimate a range of escape velocity values for Chiron by utilizing the density estimates of the two known Neptune-crossing Centaur binaries: Ceto/Phorcys ($\rho=1.4$ g cm$^{-3}$; \citealt{2007Icar..191..286G}) and Typhon/Echidna ($\rho=0.44$ g cm$^{-3}$; \citealt{2008Icar..197..260G}). This range of values is also consistent with the upper limit of Chiron's density of $\rho < 1.0$ g cm$^{-3}$ as inferred by \citet{1997AJ....113..844M}. Figure \ref{ChironGrainVelocities} shows the escape velocity range of Chiron along with the value estimated by \citet{2023A&A...676A..72B} with respect to measured dust grain velocities from outbursts by the Jupiter-family comet 67P/Churyumov-Gerasimenko \citep[]{2016MNRAS.462S.220G,2017MNRAS.469S.606A,2018MNRAS.481.1235R} and the Centaur 29P/Schwassman-Wachmann 1 \citep[]{2016Icar..272..387M,2020AJ....159..136W}, which we use as a proxy for grain velocities on Chiron and which we list in Table \ref{CometDustGrainVelocitiesTable}. 

As seen in Figure \ref{ChironGrainVelocities}, most of the outbursts from comet 67P/Churyumov-Gerasimenko have dust grain velocities that do not exceed our estimated range of values for Chiron's escape velocity. While most of the grain velocities from the Centaur 29P/Schwassman-Wachmann 1 are much greater than Chiron's estimated escape velocity, the range of values from \citet{2020AJ....159..136W} nevertheless overlaps with our range of estimated escape velocity values, including that of the shape model and density as reported by \citet{2023A&A...676A..72B}. Assuming these dust grain velocities to be representative of cometary activity on Chiron, this first-order comparison shows an entirely bound coma is a plausible outcome of a cometary outburst. In this situation, any ejecta that remains bound to Chiron's nucleus may act to augment the ring system around the Centaur, and evidence of this activity may be detectable from differences in detected structures around Chiron in (future) occultation studies \citep{2023arXiv230803458O}. However, without accurate values of its nuclear escape velocity and dust grain velocities from its cometary outbursts, we cannot rule out the possibility that Chiron is exhibiting a spatially-extended coma which is too faint to be observed at aphelion.

\rcom{Chiron's heliocentric distance of 18.9 au at the time of the first observation of the 2021-2022 observing season places it 
far beyond the boundary of direct sublimation of surface water-ice at ${\sim}3$ au \citep[]{2004come.book..317M,2017PASP..129c1001W}, therefore ruling out this mechanism as being responsible for Chiron's 2021 epoch of cometary activity. Beyond this boundary, cometary activity in the Solar System is thought to be dominated by the sublimation of the volatile species CO and CO$_{2}$ \citep[]{2012ApJ...758...29A,2012ApJ...752...15O,2013Icar..226..777R,2015SSRv..197....9C,2015ApJ...814...85B,2017PASP..129c1001W} and the transition of water ice from an amorphous to crystalline configuration \citep[]{2009AJ....137.4296J,2012AJ....144...97G,2020AJ....159..209L}. Although Chiron resides at distances too close to the Sun for these species to be maintained on its surface \citep{2009AJ....137.4296J}, sub-surface pockets of these species could persist across the Centaur's lifetime before being exposed to insolation via surface disruption \citep{1992ApJ...388..196P}. This is further corroborated by the findings of \citet{2024Icar..41316027B} that imply large quantities of CO could be preserved in the interiors of small KBOs.  An alternative mechanism proposed to explain activity throughout the Centaur population is the release of volatiles from the transition of sub-surface water ice from an amorphous to crystalline configuration caused by the propagation of the thermal wave through the object's interior \citep{2009AJ....137.4296J}. \citet{2012AJ....144...97G} found this transition can occur for Centaurs at heliocentric distances up to $16$ au. 
Though Chiron resided beyond this outer boundary during its 2021 Brightening Event, we cannot rule out Chiron being active in the years preceding the outburst, as evidenced from a possible detection of a tail-like feature by \citet{2018MNRAS.475.2512C}. Combined with the considerably different size of Chiron (volume equivalent radius 98 km; \citet{2023A&A...676A..72B}) compared to the Centaur model of \citet{2012AJ....144...97G} (50 km), this means we cannot rule out the amorphous-to-crystalline transition of water ice as being responsible for Chiron's 2021 Brightening Event.
We therefore consider newly-exposed pockets of volatile ices and the amorphous-to-crystalline water ice transition to both be viable mechanisms for Chiron's 2021 Brightening Event.} 


\begin{deluxetable*}{ccccc} \label{CometDustGrainVelocitiesTable}
\tablecaption{Comet 67P and 29P Dust Grain Velocities Used in this Analysis.}
\tablecolumns{7}
\tablehead{
\colhead{Object} & \colhead{Date} & \colhead{Grain Velocity} & \colhead{Source}\\
\colhead{} & \colhead{} & \colhead{(m/s)} & \colhead{}
}
\startdata
67P&2015 September 13 (Outburst B)&$30.2\pm1.4$&Rinaldi et al. (2018)\\
67P&2015 September 13 (Outburst B)&$22.2\pm2.2$&Rinaldi et al. (2018)\\
67P&2015 September 13 (Outburst C)&$38.4\pm2.0$&Rinaldi et al. (2018)\\
67P&2015 September 13 (Outburst D)&$64.9\pm10.0$&Rinaldi et al. (2018)\\
67P&2015 September 14 (Outburst F)&$25.30\pm1.65$&Rinaldi et al. (2018)\\
67P&2016 February 19&$24.0$&Grun et al. (2016)\\
67P&2016 July 3&$25.0\pm10.0$&Agarwal et al. (2017)\\
67P&2016 July 3&$0.41\pm0.05$&Agarwal et al. (2017)\\
\hline
29P&&$50.00 - 150.00$&Wierzchos and Womack (2020)\\
29P&2009 February 23&$154\pm18$&Miles (2016)\\
\enddata
\end{deluxetable*}

\section{Conclusions} \label{Conclusions}

We have analysed a ${\sim}1$ mag brightening in reduced/apparent magnitude by the large Centaur Chiron which started in 2021, referred to in this study as the 2021 Brightening Event. Combining multi-filter observations from multiple instruments,
we find that from 2021 June to 2023 February, Chiron has remained brighter than its pre-2021 Brightening Event magnitude, and has not exhibited any significant change in color index across time. The ${\sim}1$ mag increase of Chiron's brightness and its continued brightening across the 2021-2022 observing season disfavours rotational modulation, solar phase angle effects, and seasonal albedo features as plausible causes for this behaviour. We find the \citet{2015A&A...576A..18O} model of Chiron's reported ring system also does not account for the observed brightening. Chiron's enhanced brightening followed by its slow dimming across its 2022-2023 observing season, combined with its duration according with that of previous cometary outbursts, means we propose that the best explanation for the 2021 Brightening Event is an epoch of either new or increased cometary activity\rcom{, caused either by the amorphous-to-crystalline water ice transition or newly-exposed pockets of subsurface volatiles.} We cannot completely rule out Chiron's ring system playing a role in its observed brightening, potentially by a collision of two small, unseen shepherd moons within the ring system, highlighting the need for further occultation observations over the coming years. We find no visible coma in any of our deep-stack Gemini images, yet brightness values of Chiron's historical comae imply the Centaur could be exhibiting coma that is too faint to be observed near aphelion or alternatively is bound to the nucleus. 

We highlight that the multi-year baseline and high cadence of the ATLAS observations allowed Chiron's  brightening to be detected and discerned via analysis of the Centaur's phase curve.
The Rubin Observatory Legacy Survey of Space and Time (LSST) by the Vera C. Rubin Observatory \citep[]{2009arXiv0912.0201L,2019arXiv190108549J,2019ApJ...873..111I,2023ApJS..266...22S} will allow a 10-year dataset of multi-filter, high-cadence observations of approximately 5 million Solar System objects including the Centaur population \citep{2022zndo...5836022G}. This will enable the discovery and analysis of potentially unseen Centaur activity via phase curve analysis for many more Centaurs.
Our analysis thus serves as an example of how such a dataset can be used to detect cometary activity exhibited by Centaurs including Chiron, itself a possible fly-by target for future space missions \citep{2019EPSC...13.2025S}. 

\begin{acknowledgements}

\section{Acknowledgements}

MES was supported by the UK Science Technology Facilities Council (STFC) grants ST/V000691/1 and ST/X001253/1. MMD was supported by the UK Science Technology Facilities Council (STFC) grant ST/V506990/1. We acknowledge travel support provided by STFC for UK participation in LSST through grant ST/N002512/1 in addition to travel support provided by the Royal Astronomical Society. LJS acknowledges support by the European Research Council (ERC) under the European Union’s Horizon 2020 research and innovation program (ERC Advanced Grant KILONOVA No. 885281). PHB acknowledges support from the DIRAC Institute in the Department of Astronomy at the University of Washington, United States. The DIRAC Institute is supported through generous gifts from the Charles and Lisa Simonyi Fund for Arts and Sciences, and the Washington Research Foundation. JM acknowledges support from the Department for the Economy (DfE) Northern Ireland postgraduate studentship scheme. MSK was supported by the NASA Solar System Observations program (80NSSC20K0673)

This research has made use of data and/or services provided by the International Astronomical Union's Minor Planet Center.

This work has made use of data from the Asteroid Terrestrial-impact Last Alert System (ATLAS) project. The Asteroid Terrestrial-impact Last Alert System (ATLAS) project is primarily funded to search for near earth asteroids through NASA grants NN12AR55G, 80NSSC18K0284, and 80NSSC18K1575; byproducts of the NEO search include images and catalogs from the survey area. This work was partially funded by Kepler/K2 grant J1944/80NSSC19K0112 and HST GO-15889, and STFC grants ST/T000198/1 and ST/S006109/1. We acknowledge travel support provided by STFC for UK participation in LSST through grant ST/S006206/1. The ATLAS science products have been made possible through the contributions of the University of Hawaii Institute for Astronomy, the Queen’s University Belfast, the Space Telescope Science Institute, the South African Astronomical Observatory, and The Millennium Institute of Astrophysics (MAS), Chile. 

This work is based on observations obtained with the Samuel Oschin Telescope 48 inch Telescope at the Palomar Observatory as part of the Zwicky Transient Facility project. Major funding has been provided by the U.S. National Science Foundation under grant No. AST-1440341 and by the ZTF partner institutions: the California Institute of Technology, the Oskar Klein Centre, the Weizmann Institute of Science, the University of Maryland, the University of Washington, Deutsches Elektronen-Synchrotron, the University of Wisconsin-Milwaukee, and the TANGO Program of the University System of Taiwan.

This work makes use of observations from the Las Cumbres Observatory global telescope network.  Observations with the LCOGT 1m were obtained as part of the LCO Outbursting Objects Key (LOOK) Project (KEY2020B-009).

This research has made use of services provided by NASA's Astrophysics Data System. 


This work has also made use of data from the European Space Agency (ESA) mission
{\it Gaia} (\url{https://www.cosmos.esa.int/gaia}), processed by the {\it Gaia}
Data Processing and Analysis Consortium (DPAC,
\url{https://www.cosmos.esa.int/web/gaia/dpac/consortium}). Funding for the DPAC
has been provided by national institutions, in particular the institutions
participating in the {\it Gaia} Multilateral Agreement. 

This work has made use of data from the TRAnsiting Planets and PlanetesImals Small Telescope South (TRAPPIST-South). TRAPPIST is a project funded by the Belgian Fonds (National) de la Recherche Scientifique (F.R.S.-FNRS) under grant PDR T.0120.21. E. Jehin is a F.R.S.-FNRS Senior Research Associate. 

This work has made use of data and services provided by the Horizons system of the Jet Propulsion Laboratory. 

This project used public archival data from the Dark Energy Survey (DES). Funding for the DES Projects has been provided by the U.S. Department of Energy, the U.S. National Science Foundation, the Ministry of Science and Education of Spain, the Science and Technology FacilitiesCouncil of the United Kingdom, the Higher Education Funding Council for England, the National Center for Supercomputing Applications at the University of Illinois at Urbana-Champaign, the Kavli Institute of Cosmological Physics at the University of Chicago, the Center for Cosmology and Astro-Particle Physics at the Ohio State University, the Mitchell Institute for Fundamental Physics and Astronomy at Texas A\&M University, Financiadora de Estudos e Projetos, Funda{\c c}{\~a}o Carlos Chagas Filho de Amparo {\`a} Pesquisa do Estado do Rio de Janeiro, Conselho Nacional de Desenvolvimento Cient{\'i}fico e Tecnol{\'o}gico and the Minist{\'e}rio da Ci{\^e}ncia, Tecnologia e Inova{\c c}{\~a}o, the Deutsche Forschungsgemeinschaft, and the Collaborating Institutions in the Dark Energy Survey. The Collaborating Institutions are Argonne National Laboratory, the University of California at Santa Cruz, the University of Cambridge, Centro de Investigaciones Energ{\'e}ticas, Medioambientales y Tecnol{\'o}gicas-Madrid, the University of Chicago, University College London, the DES-Brazil Consortium, the University of Edinburgh, the Eidgen{\"o}ssische Technische Hochschule (ETH) Z{\"u}rich, Fermi National Accelerator Laboratory, the University of Illinois at Urbana-Champaign, the Institut de Ci{\`e}ncies de l'Espai (IEEC/CSIC), the Institut de F{\'i}sica d'Altes Energies, Lawrence Berkeley National Laboratory, the Ludwig-Maximilians Universit{\"a}t M{\"u}nchen and the associated Excellence Cluster Universe, the University of Michigan, the National Optical Astronomy Observatory, the University of Nottingham, the Ohio State University, the OzDES Membership Consortium, the University of Pennsylvania, the University of Portsmouth, SLAC National Accelerator Laboratory, Stanford University, the University of Sussex, and Texas A\&M University. Based in part on observations at Cerro Tololo Inter-American Observatory, National Optical Astronomy Observatory, which is operated by the Association of Universities for Research in Astronomy (AURA) under a cooperative agreement with the National Science Foundation. Database access and other data services are provided by the Astro Data Lab.

This work has made use of data and services from the Panoramic Survey Telescope and Rapid Response System (Pan-STARRS). The Pan-STARRS1 Surveys (PS1) and the PS1 public science archive have been made possible through contributions by the Institute for Astronomy, the University of Hawaii, the Pan-STARRS Project Office, the Max-Planck Society and its participating institutes, the Max Planck Institute for Astronomy, Heidelberg and the Max Planck Institute for Extraterrestrial Physics, Garching, The Johns Hopkins University, Durham University, the University of Edinburgh, the Queen's University Belfast, the Harvard-Smithsonian Center for Astrophysics, the Las Cumbres Observatory Global Telescope Network Incorporated, the National Central University of Taiwan, the Space Telescope Science Institute, the National Aeronautics and Space Administration under Grant No. NNX08AR22G issued through the Planetary Science Division of the NASA Science Mission Directorate, the National Science Foundation Grant No. AST-1238877, the University of Maryland, Eotvos Lorand University (ELTE), the Los Alamos National Laboratory, and the Gordon and Betty Moore Foundation. Data from Pan-STARRS were obtained from the MAST data archive at the Space Telescope Science Institute. 

This research utilized data and services provided by the Canadian Astronomy Data Center (CADC) Solar System Object Image Search.
This research utilized data and services provided by the Catalog Archive Server Jobs System, developed by the Johns Hopkins University/Sloan Digital Sky Survey (JHU/SDSS) team. 


This project has received funding from the European Union’s Horizon 2020 research and innovation programme under the Marie Sk{\l}odowska-Curie grant agreement No. 101032479.

This research made use of Photutils, an Astropy package for detection and photometry of astronomical sources \citep{larry_bradley_2023_7946442}.

All photometric data used in this study is provided in full as supplementary information accompanying this paper, and are located in the appendix section of this paper.  Raw and Calibrated Observations from Las Cumbres Observatory used in this study are available at the LCO Science Archive (\url{https://archive.lco.global}; proposal code KEY2020B-009) after an embargo/proprietary period of 12 months. 

We would like to thank Jose-Luiz Ortiz 
for their useful discussions about the Chiron ring model. 

\rcom{The authors thank the anonymous reviewers for their comments and feedback that improved this paper}

\facilities{ATLAS (Chile, Haleakala, Mauna Loa, and South Africa telescopes), PO:1.2m (ZTF), PS1, Blanco (DES), Gaia, LCOGT (1-m telescopes), TRAPPIST, Gemini Gillett.}

\end{acknowledgements}

\software{Astropy 
\citep[]{2013A&A...558A..33A,2018AJ....156..123A},
BANZAI \citep{2018SPIE10707E..0KM},
\calviacat \citep{2021zndo...5061298K},
ccdproc \citep{craig-2017},
DRAGONS \citep[]{2019ASPC..523..321L,Labrie_2023},
Jupyter Notebook \citep{soton403913},
math \citep{van1995python},
Matplotlib \citep{Hunter:2007},
NEOExchange \citep{2021Icar..36414387L},
Numpy \citep{2011CSE....13b..22V,harris2020array},
os \citep{van1995python},
Pandas \citep{reback2020pandas},
PHOTOMETRYPIPELINE \citep{2017A&C....18...47M},
Photutils \citep{larry_bradley_2023_7946442},
python (\url{https://www.python.org}),
SAOImageDS9 \citep{2019zndo...2530958J},
SciPy \citep{2020NatMe..17..261V},
Source Extractor Python \citep[]{1996A&AS..117..393B,2016JOSS....1...58B}}

\appendix 

\section{Observations} \label{Appendix}

All observations used in this study are listed in the tables below, separated by telescope/survey.

\restartappendixnumbering

\begin{deluxetable*}{ccccccc} \label{ATLASTable}
\tablecaption{ATLAS Observations of Chiron}
\tablecolumns{7}
\tablehead{
\colhead{MJD\tablenotemark{a}} & \colhead{Magnitude} & \colhead{Magnitude} & \colhead{Filter} & \colhead{$R_{H}$\tablenotemark{b}} & \colhead{$\Delta$\tablenotemark{c}} & \colhead{$\alpha$\tablenotemark{d}} \\
\colhead{} & \colhead{} & \colhead{Uncertainty} & \colhead{} & \colhead{} & \colhead{} & \colhead{}\\
\colhead{} & \colhead{} & \colhead{} & \colhead{} & \colhead{(au)} & \colhead{(au)} & \colhead{(deg)}
}
\startdata
57229.491001 & 17.945 & 0.141 & $o$ & 18.150299 & 17.456002 & 2.3875 \\
57229.493189 & 17.784 & 0.128 & $o$ & 18.150300 & 17.455976 & 2.3874 \\
57229.514768 & 17.899 & 0.083 & $o$ & 18.150315 & 17.455723 & 2.3866 \\
57229.541811 & 18.083 & 0.097 & $o$ & 18.150333 & 17.455407 & 2.3855 \\
57246.461081 & 18.565 & 0.103 & $c$ & 18.161646 & 17.289222 & 1.6685 \\
57246.466938 & 18.243 & 0.093 & $c$ & 18.161649 & 17.289175 & 1.6682 \\
57246.502626 & 18.597 & 0.122 & $c$ & 18.161673 & 17.288894 & 1.6665 \\
57313.399531 & 18.176 & 0.131 & $c$ & 18.205515 & 17.402518 & 1.8983 \\
57330.396038 & 18.073 & 0.153 & $o$ & 18.216428 & 17.623558 & 2.5419 \\
57356.286485 & 18.316 & 0.137 & $o$ & 18.232879 & 18.049571 & 3.0602 \\
\enddata
\tablecomments{This table is published in its entirety in the machine-readable format. A portion is shown here for guidance for regarding its form and content.}
\tablecomments{All observation exposure times are 30 seconds.}
\tablenotetext{a}{Modified Julian date of observation, measured at midpoint of exposure.}
\tablenotetext{b}{Heliocentric distance}
\tablenotetext{c}{Geocentric distance}
\tablenotetext{d}{Solar phase angle}
\end{deluxetable*}

\begin{deluxetable*}{ccccccc} \label{ZTFTable}
\tablecaption{ZTF Observations of Chiron}
\tablecolumns{7}
\tablehead{
\colhead{MJD\tablenotemark{a}} & \colhead{Magnitude} & \colhead{Magnitude} & \colhead{Filter} & \colhead{$R_{H}$\tablenotemark{b}} & \colhead{$\Delta$\tablenotemark{c}} & \colhead{$\alpha$\tablenotemark{d}} \\
\colhead{} & \colhead{} & \colhead{Uncertainty} & \colhead{} & \colhead{} & \colhead{} & \colhead{}\\
\colhead{} & \colhead{} & \colhead{} & \colhead{} & \colhead{(au)} & \colhead{(au)} & \colhead{(deg)}
}
\startdata
58789.206690 &18.739 & 0.098 &$ZTF-g$ & 18.816841 & 18.019101 & 1.8359 \\
58789.229294 &18.311 & 0.061 &$ZTF-r$ & 18.816846 & 18.019345 & 1.8369 \\
58828.127928 &18.363 & 0.089 &$ZTF-r$ & 18.823712 & 18.583454 & 2.9246 \\
58831.071377 &18.369 & 0.065 &$ZTF-r$ & 18.824212 & 18.633899 & 2.9539 \\
58855.191424 &18.568 & 0.106 &$ZTF-r$ & 18.828206 & 19.051763 & 2.8959 \\
58863.091262 &18.463 & 0.073 &$ZTF-r$ & 18.829474 & 19.182602 & 2.7667 \\
58863.121482 &18.953 & 0.118 &$ZTF-g$ & 18.829479 & 19.183090 & 2.7661 \\
59034.382199 &18.407 & 0.094 &$ZTF-r$ & 18.852102 & 18.763725 & 3.0869 \\
59038.378009 &18.424 & 0.105 &$ZTF-r$ & 18.852518 & 18.697623 & 3.0678 \\
59044.358461 &18.774 & 0.120 &$ZTF-g$ & 18.853133 & 18.599799 & 3.0134 \\
\enddata
\tablecomments{This table is published in its entirety in the machine-readable format. A portion is shown here for guidance for regarding its form and content.}
\tablecomments{All observation exposure times are 30 seconds.}
\tablenotetext{a}{Modified Julian date of observation}
\tablenotetext{b}{Heliocentric distance}
\tablenotetext{c}{Geocentric distance}
\tablenotetext{d}{Solar phase angle}
\end{deluxetable*}

\begin{deluxetable*}{cccccccc} \label{PanSTARRSTable}
\tablecaption{Pan-STARRS Observations of Chiron}
\tablecolumns{7}
\tablehead{
\colhead{MJD\tablenotemark{a}} & \colhead{Magnitude} & \colhead{Magnitude} & \colhead{Filter} & \colhead{Exposure} & \colhead{$R_{H}$\tablenotemark{b}} & \colhead{$\Delta$\tablenotemark{c}} & \colhead{$\alpha$\tablenotemark{d}} \\
\colhead{} & \colhead{} & \colhead{Uncertainty} & \colhead{} & \colhead{Time} & \colhead{} & \colhead{} & \colhead{}\\
\colhead{} & \colhead{} & \colhead{} & \colhead{} & \colhead{(s)} & \colhead{(au)} & \colhead{(au)} & \colhead{(deg)}\\
}
\startdata
54992.603300&17.5890&0.0324&$y$&30.0&15.852166&15.419669&3.3679\\
54992.615072&17.6314&0.0372&$y$&30.0&15.852182&15.419506&3.3676\\
55345.600003&17.7551&0.0448&$y$&30.0&16.322017&16.165605&3.5340\\
55345.612039&17.6574&0.0477&$y$&30.0&16.322033&16.165422&3.5339\\
55369.612051&17.6963&0.0205&$z$&30.0&16.352523&15.819741&3.0845\\
55439.437320&17.6177&0.0086&$r$&40.0&16.440162&15.450846&0.7197\\
55439.448968&17.6046&0.0084&$r$&40.0&16.440177&15.450896&0.7203\\
55441.438908&17.5285&0.0075&$i$&45.0&16.442651&15.460135&0.8237\\
55444.289450&18.0477&0.0106&$g$&43.0&16.446193&15.475389&0.9771\\
55444.301023&18.0487&0.0108&$g$&43.0&16.446207&15.475455&0.9777\\
\enddata
\tablecomments{This table is published in its entirety in the machine-readable format. A portion is shown here for guidance for regarding its form and content.}
\tablenotetext{a}{Modified Julian date of observation, measured at midpoint of exposure}
\tablenotetext{b}{Heliocentric distance}
\tablenotetext{c}{Geocentric distance}
\tablenotetext{d}{Solar phase angle}
\end{deluxetable*}

\begin{deluxetable*}{ccccccc} \label{DESTable}
\tablecaption{DES Observations of Chiron}
\tablecolumns{7}
\tablehead{
\colhead{MJD\tablenotemark{a}} & \colhead{Magnitude} & \colhead{Magnitude} & \colhead{Filter} & \colhead{$R_{H}$\tablenotemark{b}} & \colhead{$\Delta$\tablenotemark{c}} & \colhead{$\alpha$\tablenotemark{d}} \\
\colhead{} & \colhead{} & \colhead{Uncertainty} & \colhead{} & \colhead{} & \colhead{} & \colhead{}\\
\colhead{} & \colhead{} & \colhead{} & \colhead{} & \colhead{(au)} & \colhead{(au)} & \colhead{(deg)}
}
\startdata
56546.247289&18.2370&0.0042&$g$&17.618302&16.624231&0.5310\\
56951.130462&18.5977&0.0047&$g$&17.951314&17.224747&2.2170\\
56958.142839&18.1324&0.0176&$Y$&17.956625&17.319614&2.4769\\
57294.162942&17.9745&0.0140&$Y$&18.193055&17.236986&0.9678\\
57327.048550&18.1877&0.0043&$i$&18.214288&17.575397&2.4321\\
57327.049921&18.2352&0.0040&$r$&18.214289&17.575416&2.4321\\
57327.051287&18.6546&0.0046&$g$&18.214290&17.575435&2.4321\\
57657.141772&18.0368&0.0042&$i$&18.408731&17.426478&0.6410\\
57657.143145&18.1057&0.0040&$r$&18.408732&17.426484&0.6410\\
57657.144516&18.5195&0.0046&$g$&18.408732&17.426490&0.6411\\
\enddata
\tablecomments{This table is published in its entirety in the machine-readable format. A portion is shown here for guidance for regarding its form and content.}
\tablenotetext{a}{Modified Julian date of observation}
\tablenotetext{b}{Heliocentric distance}
\tablenotetext{c}{Geocentric distance}
\tablenotetext{d}{Solar phase angle}
\end{deluxetable*}

\begin{deluxetable*}{ccccccc} \label{GaiaTable}
\tablecaption{Gaia Observations of Chiron}
\tablecolumns{7}
\tablehead{
\colhead{MJD\tablenotemark{a}} & \colhead{Magnitude} & \colhead{Magnitude} & \colhead{Filter} & \colhead{$R_{H}$\tablenotemark{b}} & \colhead{$\Delta$\tablenotemark{c}} & \colhead{$\alpha$\tablenotemark{d}} \\
\colhead{} & \colhead{} & \colhead{Uncertainty} & \colhead{} & \colhead{} & \colhead{} & \colhead{}\\
\colhead{} & \colhead{} & \colhead{} & \colhead{} & \colhead{(au)} & \colhead{(au)} & \colhead{(deg)}
}
\startdata
56961.329590&18.4157&0.0077&$Gaia-G$&17.959033&17.365889&2.5822\\
56961.329646&18.4157&0.0077&$Gaia-G$&17.959033&17.365890&2.5823\\
56961.329702&18.4157&0.0077&$Gaia-G$&17.959033&17.365890&2.5823\\
56961.329759&18.4157&0.0077&$Gaia-G$&17.959033&17.365891&2.5823\\
56961.329815&18.4157&0.0077&$Gaia-G$&17.959034&17.365892&2.5823\\
56961.329871&18.4157&0.0077&$Gaia-G$&17.959034&17.365893&2.5823\\
56961.329927&18.4157&0.0077&$Gaia-G$&17.959034&17.365894&2.5823\\
56961.329984&18.4157&0.0077&$Gaia-G$&17.959034&17.365895&2.5823\\
56961.330040&18.4157&0.0077&$Gaia-G$&17.959034&17.365896&2.5823\\
56961.403597&18.4020&0.0075&$Gaia-G$&17.959089&17.366985&2.5846\\
\enddata
\tablecomments{This table is published in its entirety in the machine-readable format. A portion is shown here for guidance for regarding its form and content.}
\tablenotetext{a}{Modified Julian date of observation}
\tablenotetext{b}{Heliocentric distance}
\tablenotetext{c}{Geocentric distance}
\tablenotetext{d}{Solar phase angle}
\end{deluxetable*}

\begin{deluxetable*}{cccccccccc} \label{LOOKTable}
\tablecaption{LOOK Observations of Chiron}
\tablecolumns{7}
\tablehead{
\colhead{MJD\tablenotemark{a}} & \colhead{$R_{H}$\tablenotemark{b}} & \colhead{$\Delta$\tablenotemark{c}} & \colhead{$\alpha$\tablenotemark{d}} & \colhead{Seeing} & \colhead{Filter} & \colhead{Exposure} & \colhead{Airmass\tablenotemark{e}} & \colhead{Magnitude} & \colhead{Magnitude}\\
\colhead{} & \colhead{} & \colhead{} & \colhead{} & \colhead{} & \colhead{} & \colhead{Time} & \colhead{} & \colhead{} & \colhead{Uncertainty}\\
\colhead{} & \colhead{(au)} & \colhead{(au)} & \colhead{(deg)} & \colhead{(arcsec)} & \colhead{} & \colhead{(seconds)} & \colhead{} & \colhead{} & \colhead{}
}
\startdata
59463.320833 & 18.867587 & 17.971398 & 1.4368 &    1.73 &      $w$ &    245.0 &    1.344 & 17.802 &  0.044 \\
59463.324306 & 18.867587 & 17.971371 & 1.4366 &    1.68 &      $w$ &    245.0 &    1.355 & 17.795 &  0.051 \\
59463.327083 & 18.867587 & 17.971350 & 1.4365 &    1.62 &      $w$ &    245.0 &    1.367 & 17.771 &  0.042 \\
59463.330556 & 18.867587 & 17.971323 & 1.4363 &    1.61 &      $w$ &    245.0 &    1.380 & 17.769 &  0.043 \\
59827.018750 & 18.834124 & 17.981445 & 1.6770 &    1.92 &     $g'$ &    180.1 &    1.347 & 18.178 &  0.033 \\
59827.020833 & 18.834124 & 17.981425 & 1.6769 &    2.06 &     $r'$ &    180.1 &    1.332 & 17.936 &  0.034 \\
59827.023611 & 18.834124 & 17.981400 & 1.6768 &    2.01 &     $g'$ &    180.1 &    1.318 & 18.241 &  0.032 \\
59827.025694 & 18.834123 & 17.981381 & 1.6767 &    1.71 &     $r'$ &    180.1 &    1.304 & 17.885 &  0.031 \\
59840.381944 & 18.832074 & 17.882598 & 1.0286 &    2.09 &     $g'$ &    180.1 &    1.136 & 18.156 &  0.022 \\
59840.386806 & 18.832073 & 17.882571 & 1.0284 &    2.10 &     $g'$ &    180.1 &    1.147 & 18.159 &  0.022 \\
\enddata
\tablecomments{This table is published in its entirety in the machine-readable format. A portion is shown here for guidance for regarding its form and content.}
\tablenotetext{a}{Modified Julian date of observation}
\tablenotetext{b}{Heliocentric distance}
\tablenotetext{c}{Geocentric distance}
\tablenotetext{d}{Solar phase angle}
\tablenotetext{e}{Mean airmass of Chiron during exposure.}
\end{deluxetable*}

\begin{deluxetable*}{ccccccc} \label{TRAPPISTSouthTable}
\tablecaption{TRAPPIST-South Observations of Chiron}
\tablecolumns{7}
\tablehead{
\colhead{MJD\tablenotemark{a}} & \colhead{Magnitude} & \colhead{Magnitude} & \colhead{Filter} & \colhead{$R_{H}$\tablenotemark{b}} & \colhead{$\Delta$\tablenotemark{c}} & \colhead{$\alpha$\tablenotemark{d}} \\
\colhead{} & \colhead{} & \colhead{Uncertainty} & \colhead{} & \colhead{} & \colhead{} & \colhead{}\\
\colhead{} & \colhead{} & \colhead{} & \colhead{} & \colhead{(au)} & \colhead{(au)} & \colhead{(deg)}
}
\startdata
56063.41221&17.960&0.055&$R$&17.152954&17.359322&3.2885\\
56063.41861&18.043&0.065&$R$&17.152960&17.359225&3.2886\\
56063.42314&18.312&0.066&$V$&17.152965&17.359156&3.2887\\
56063.42612&18.999&0.110&$B$&17.152968&17.359111&3.2887\\
56063.42901&17.976&0.064&$R$&17.152971&17.359067&3.2887\\
56063.43192&18.279&0.077&$V$&17.152974&17.359023&3.2888\\
56084.40271&18.336&0.111&$V$&17.174743&17.033325&3.3678\\
56084.41155&18.732&0.152&$B$&17.174753&17.033188&3.3678\\
56101.39035&17.866&0.050&$R$&17.192274&16.777463&3.1319\\
56101.39330&17.854&0.053&$R$&17.192277&16.777420&3.1319\\
\enddata
\tablecomments{This table is published in its entirety in the machine-readable format. A portion is shown here for guidance for regarding its form and content.}
\tablenotetext{a}{Modified Julian date of observation}
\tablenotetext{b}{Heliocentric distance}
\tablenotetext{c}{Geocentric distance}
\tablenotetext{d}{Solar phase angle}
\end{deluxetable*}

\begin{deluxetable*}{ccccccccccc} \label{tab:LLP-obs}
\tablecaption{Gemini Observing Details and Photometry}
\tablecolumns{10}
\tablewidth{0pt}
\tablehead{
\colhead{MJD\tablenotemark{a}} &
\colhead{$R_H$\tablenotemark{b}} &
\colhead{$\Delta$\tablenotemark{c}} &
\colhead{$\alpha$\tablenotemark{d}} &
\colhead{Seeing} &
\colhead{Filter} &
\colhead{Total Exp.} &
\colhead{Airmass\tablenotemark{e}} &
\colhead{Magnitude} &
\colhead{Magnitude}\\      
\colhead{Date} & 
\colhead{} &
\colhead{} &
\colhead{} &
\colhead{} & 
\colhead{} & 
\colhead{Time} &
\colhead{} &
\colhead{} &
\colhead{Uncertainty}\\
\colhead{} & 
\colhead{(au)} &
\colhead{(au)} &
\colhead{(deg)} &
\colhead{(arcsec)} & 
\colhead{} & 
\colhead{(seconds)} &
\colhead{} &
\colhead{} &
\colhead{}\\
}

\startdata
59446.57370 & 18.868 & 18.135 & 2.166 & 0.945 &g &180 &1.040 & 18.327 & 0.039 \\
59446.57704 & 18.868 & 18.135 & 2.166 & 0.898 &r &360 &1.046 & 17.897 & 0.027 \\
59446.58361 & 18.868 & 18.135 & 2.165 & 0.847 &i &180 &1.053 & 17.784 & 0.030 \\
59521.38615 & 18.865 & 18.016 & 1.587 & 1.216 &g &990 &1.109 & 18.012 & 0.021 \\
59521.40480 & 18.865 & 18.016 & 1.588 & 0.966 &i &990 &1.163 & 17.451 & 0.006 \\
59557.31971 & 18.863 & 18.487 & 2.791 & 1.488 &g &990 &1.232 & 18.185 & 0.019 \\
59557.33648 & 18.863 & 18.488 & 2.791 & 1.151 &i &990 &1.329 & 17.712 & 0.019 \\
59754.58140 & 18.844 & 19.049 & 3.011 & 0.492 &g &360 &1.460 & 18.601 & 0.046 \\
59754.58683 & 18.844 & 19.048 & 3.011 & 0.422 &r &360 &1.413 & 18.100 & 0.035 \\
59793.50172 & 18.839 & 18.406 & 2.828 & 1.040 &g &900 &1.247 & 18.483 & 0.040 \\
59793.51530 & 18.839 & 18.406 & 2.828 & 0.909 &r &600 &1.190 & 17.932 & 0.033 \\
59793.52565 & 18.839 & 18.406 & 2.827 & 0.808 &i &600 &1.149 & 17.787 & 0.029 \\
\enddata
\tablenotetext{a}{UTC at start of image sequence.}
\tablenotetext{b}{Heliocentric distance}
\tablenotetext{c}{Geocentric distance}
\tablenotetext{d}{Solar phase angle}
\tablenotetext{e}{Mean airmass of Chiron during exposure.}
\end{deluxetable*}

\bibliography{bibliography.bib}{}
\bibliographystyle{aasjournal}



\end{document}